\documentclass[aps,11pt,tightenlines,amsmath,amssymb,showpacs]{revtex4}
\usepackage{graphicx}
\usepackage{dcolumn}
\usepackage{bm}

\begin{document}

\title{Evidence for an axion-like particle from PKS 1222+216?}

\author{Fabrizio Tavecchio}
\email{fabrizio.tavecchio@brera.inaf.it}
\affiliation{INAF -- Osservatorio Astronomico di Brera, Via E. Bianchi 46, I--23807 Merate, Italy}
\author{Marco Roncadelli}
\email{marco.roncadelli@pv.infn.it}
\affiliation{INFN, Sezione di Pavia, Via A. Bassi 6, I -- 27100 Pavia, Italy}
\author{Giorgio Galanti}
\email{giorgio_galanti@libero.it}
\affiliation{Dipartimento di Fisica, Universit\`a dell'Insubria, Via Valleggio 11, I -- 22100 Como, Italy}
\author{Giacomo Bonnoli}
\email{giacomo.bonnoli@brera.inaf.it}
\affiliation{INAF -- Osservatorio Astronomico di Brera, Via E. Bianchi 46, I--23807 Merate, Italy~}

\begin{abstract}
The surprising discovery by MAGIC of an intense, rapidly varying emission in the energy range $70 - 400 \, {\rm GeV}$ from the flat spectrum radio quasar PKS 1222+216 represents a challenge for all interpretative scenarios. Indeed, in order to avoid absorption of 
$\gamma$ rays in the dense ultraviolet radiation field of the broad line region (BLR), one is forced to invoke some unconventional astrophysical picture, like for instance the existence of a very compact ($r\sim 10^{14}$ cm) emitting blob at a large distance ($R \sim10^{18}$ cm) from the jet base. We offer the investigation of a scenario based on the standard blazar model for PKS 1222+216 where $\gamma$ rays are produced close to the central engine, but we add the new assumption that inside the source photons can oscillate into axion-like particles (ALPs), which are a generic prediction of several extensions of the Standard Model of elementary particle interactions. As a result, a considerable fraction of very-high-energy photons can escape absorption from the BLR through the mechanism of photon-ALP oscillations much in the same way as they largely avoid absorption from extragalactic background light when propagating over cosmic distances in the presence of large-scale magnetic fields in the nG range. {\it In addition} we show that the above MAGIC observations and the simultaneous {\it Fermi}/LAT observations in the energy range $0.3 - 3\, {\rm GeV}$ can {\it both} be explained by a standard spectral energy distribution for experimentally allowed values of the model parameters. In particular, we need a very light ALP just like in the case of photon-ALP oscillations in cosmic space. Moreover, we find it quite tantalizing that the most favorable value of the photon-ALP coupling happens to be the same in both situations. Although our ALPs cannot contribute to the cold dark matter, they are a viable candidate for the quintessential dark energy. An astrophysical test of our scenario is proposed and an independent laboratory check for the existence of an ALP with the properties required by our picture  will be performed with the planned upgrade of the photon regeneration experiment ALPS at DESY and with the next generation of solar axion detectors like IAXO.
\end{abstract}

\pacs{14.80.Va, 95.30.--k, 95.85.Pw, 95.85.Ry}

\maketitle


\section{INTRODUCTION}

Over the last decade, growing astrophysical interest has been attracted by axion-like-particles (ALPs)~\cite{csachi1,csaki2,peloso,pvlas,massotoldra,fair,bmr2010,hooperserpico,dmr,darma,dgr,simetetal2008,prada2009,pradabis,horns}. They are very light pseudo-scalar bosons $a$ characterized by a two-photon coupling $a \gamma \gamma$, which are a generic prediction of several extensions of the Standard Model (SM) such as four-dimensional ordinary and supersymmetric models~\cite{susy}, Kaluza-Klein theories~\cite{kaluzaklein} and especially superstring theories~\cite{witten,cicoli} (for a review, see~\cite{alprev}). In the presence of an external magnetic field, the $a \gamma \gamma$ coupling produces a mismatch between the interaction eigenstates and the propagation eigenstates, thereby giving rise to the phenomena of single photon-ALP conversion $\gamma \to a$ ($a \to \gamma$) and double photon-ALP conversion $\gamma \to a \to \gamma$ ($a \to \gamma \to a$), the latter being a photon-ALP oscillation, quite similar to the oscillation involving two massive neutrinos of different flavour (apart from the fact that the external field is needed to compensate for the photon and ALP spin difference)~\cite{sikivie,raffeltstodolsky1988}. 

As far as astrophysics is concerned, their most striking feature is to drastically alter photon propagation over cosmic distances and at very high energies (VHEs), namely for $E > 50 \, {\rm GeV}$. As a consequence, blazars are the natural astronomical sources to look for photon-ALP conversion effects, even because the currently operating Imaging Atmospheric Cherenkov Telescopes (IACTs) like H.E.S.S.~\cite{hess}, MAGIC~\cite{magic} and VERITAS~\cite{veritas} provide excellent data in the VHE range up to $\sim 10 \, {\rm TeV}$ and the planned detectors like CTA~\cite{cta}, HAWC~\cite{hawc} and HiSCORE~\cite{hiscore} will be able to explore the VHE range up to $100 \, {\rm TeV}$ with much better sensitivity (HiSCORE is supposed to reach PeV energies).  

It goes without saying that various scenarios have been proposed. A possibility is that a single photon-ALP conversion $\gamma \to a$ occurs either inside the blazar~\cite{hooperserpico} or in extragalactic space~\cite{dmr} where a large-scale magnetic field in the nG range is supposed to exist, which  is consistent with presently available upper bounds~\cite{lsmf} and AUGER results~\cite{auger}. In either case, the result is a dimming of the source above a characteristic energy $E_*$ which depends on the model parameters, and so it shows up as a feature somewhere in the observed $\gamma$-ray spectrum. A different option -- called DARMA scenario (achronym of De Angelis, Roncadelli and Mansutti) -- contemplates photon-ALP oscillations $\gamma \to a \to \gamma$ as taking place in extragalactic space~\cite{darma,dgr} provided that a large-scale magnetic field in the nG range is present. Alternatively, the conversion $\gamma \to a$ can occur within the blazar whereas the re-conversion $a \to \gamma$ is supposed to happen in the Milky-Way~\cite{simetetal2008}. Needless to say, also both cases can be realized at once~\cite{prada2009}. 

Actually, photon-ALP oscillations are particularly intriguing in view of our later considerations, since what we are used to simply regard as a {\it photon} behaves for some time as a ``true photon'' and for some time as an ALP. Now, ``true photons'' can disappear from the beam along their way to us through the $\gamma \gamma \to e^+ e^-$ scattering~\cite{pairproduction1,pairproduction2} with low-energy infrared/optical/ultraviolet photons of the extragalactic background light (EBL), namely the diffuse radiation produced by stars during the whole cosmic evolution (for a review, see~\cite{ebl}). Yet, ALPs propagate totally unaffected by the EBL, since the process $a \gamma \to \gamma$ is kinematically forbidden and the $a \gamma \to e^+ e^-$ scattering has a ridiculously small cross-section. As a consequence, owing to photon-ALP oscillations the effective photon optical depth $\tau_{\rm eff} (E)$ gets smaller~\cite{commentA}, and even a tiny decrease entails a large enhancement of the photon survival probability $P_{\gamma \to \gamma} (E)$ as compared with conventional physics since the two quantities are related by 
\begin{equation}
\label{tM}
P_{\gamma \to \gamma} (E) = e^{- \tau_{\rm eff} (E)}~.
\end{equation}
In addition, because the EBL dimming increases with $E$ (see e.g. Fig. \ref{fig:france}) whereas the photon-ALP oscillations alone are $E$-independent, the resulting observed blazar spectra become harder than expected and in particular the ``$\gamma$-ray horizon" gets considerably enlarged (a very thorough analysis of this point is contained in~\cite{dgr}). To date no clear-cut evidence for ALPs exists but some suggestions pointing toward their existence have been put forward in~\cite{pradabis} and especially in~\cite{horns}, although other (more conventional) astrophysical solutions have been proposed~\cite{essey}.

Coming back to blazars, they dominate the extragalactic $\gamma$-ray sky, both at high energy ($>100$ MeV) and at VHE. Their powerful non-thermal emission, spanning the entire electromagnetic spectrum, is produced in a relativistic jet pointing toward the Earth. Their spectral energy distribution (SED) shows two well defined ``humps''. The first one -- peaking somewhere between the IR and the X-ray bands -- derives from the synchrotron emission of relativistic electrons (or, more generally, e$^+ e^-$ pairs) in the jet. The origin of the second component which exhibits a maximum at $\gamma$-ray energies is more debated. Leptonic models~\cite{leptonic} attribute it to the inverse Compton emission of the same electrons responsible for the lower energy bump (with the possible additional contribution from external photons). Hadronic models (for a review, see~\cite{hadronic}), instead, assume that the $\gamma$ rays are the leftover of reactions involving relativistic hadrons. 

Blazars are further divided into two broad groups, BL Lac objects and flat spectrum radio quasars (FSRQs)~\cite{urry}. BL Lacs are defined by the weakness (or even absence) of thermal features (most notably broad emission lines) in their optical spectra. This evidence leads to the common belief that the nuclear region of BL Lacs, where the jet forms and accelerates, is rather poor of soft photons. On the contrary, FSRQs display luminous broad ($v>1000$ km s$^{-1}$) emission lines, indicating the existence of photo-ionized clouds rapidly rotating around the central black hole and organized in the so-called broad line region (BLR) (see 
e.g.~\cite{netzer2008}). 

Besides their importance for the study of the structure and functioning of relativistic jets, growing interest for blazars is motivated by the use of their intense $\gamma$-ray beam as a probe of the EBL (see e.g.~\cite{aharonianetal2006}) and of the large-scale  magnetic fields (see e.g.~\cite{neronovvovk2010}), and even more fundamentally for the study of new physical phenomena beyond the SM -- like indeed ALPs -- along with quite radical departures from conventional physics such as violation of Lorentz invariance (for a review, see~\cite{liberatimaccione2009}). 

As far as blazar observations are concerned, the recent evidence of VHE emission by FSRQs poses a quite serious challenge. As stressed above, the surrounding of the inner jet in FSRQs is rich of optical/ultraviolet photons emitted by the BLR, necessarily implying a huge optical depth for $\gamma$ rays above 10 -- 20 GeV (see e.g.~\cite{liubai2002, tavecchiomazin2009}). Therefore, the observation of some FSRQs at TeV energies raised great surprise~\cite{albertetal2008, aleksicetal2011a, wagnerbehera2010}. Moreover, the detection of an intense VHE emission from PKS 1222+216 at redshift $z = 0.432$ in the energy range 70 -- 400 GeV~\cite{aleksicetal2011a} observed by MAGIC to double its flux in only about 10 minutes -- thereby flagging the extreme compactness of the emitting region -- is very difficult to fit within the standard blazar models~\cite{tavecchioetal2011}. 

So far, the only possibility to solve the apparent contradiction with conventional expectations arising from both the detection of the intense and relatively hard VHE flux and the rapid variation appears to invoke the existence of very small ($r \sim10^{14}$ cm) emitting regions (or beams of particles) beyond the BLR ($R\sim 10^{18}$ cm), that is at a large distance from the central engine~\cite{tavecchioetal2011,nalewajko2012,dermer}. 
 
An additional question is that PKS 1222+216 has also been simultaneously detected by {\it Fermi}/LAT in the energy range $0.3 - 3\, {\rm GeV}$~\cite{fermi}. 

So, one would like not only to understand why the VHE $\gamma$ rays are actually {\it emitted} by PKS 1222+216 but also to find a {\it realistc and physically motivated} SED that fits {\it both} the observed {\it Fermi}/LAT and the MAGIC spectra, which is a logically distinct and more ambitious task.

Our proposal to naturally explain all observations assumes the validity of a standard blazar model for photon production, but we add the new assumption that photon-ALP oscillations take place inside the source in such a way that a considerable fraction of VHE photons can leave it, indeed much in the same fashion that a sizable amount of VHE photons emitted by blazars largely avoid EBL absorption if extragalactic space is permeated by a large-scale magnetic field in the nG range. More specifically, we envisage that the $\gamma \to a$ conversion occurs before most of the VHE photons reach the BLR. Accordingly, ALPs traverse this region unimpeded and outside it the re-conversion $a \to \gamma$ takes place either in the same magnetic field of the source or in that of the host galaxy. Thus, our proposal differs from any previously considered one. Moreover, we find that for observationally allowed values of the parameters of our model the resulting SED looks quite realistic and nicely fits the {\it Fermi}/LAT and MAGIC spectra observed at the same time. 
 
Our aim is to carry out an investigation of the proposed scenario. The plan of the paper is as follows. The mechanism of photon-ALP oscillations is reviewed in general terms in Section II. Our model for PKS 1222+216 is described in Section III and its results are presented in Section IV. In Section V we provide a specific picture which gives rise to the required SED, structurally similar to the model developed in~\cite{tavecchioetal2011} but with an important difference. Finally, in Section VI we draw our conclusions, comparing our proposal to others recently put forward, stressing the possibility of an astrophysical check of our model and emphasizing that an ALP having precisely the right properties needed for our scenario to work will be searched for in the near future in two distinct laboratory experiments.

\section{PHOTON-ALP OSCILLATIONS}

As we said, a generic feature of several extensions of the SM is the prediction of axion-like particles (ALPs). As the name indeed suggests, they are a generalization of the axion (for a review, see~\cite{axionrev}) -- the pseudo-Goldstone boson arising from the breakdown of the global Peccei-Quinn symmetry $U(1)_{\rm PQ}$ proposed as a natural solution to the strong CP problem -- but important differences exist between the axion and ALPs due to the fact that the axion arises within a very specific framework whereas in dealing with ALPs the aim is to bring out their properties as much as possible in a model-independent manner~\cite{alprev}. Because of this different attitude, two main differences come about. One is that only the ALP-photon interaction is taken into account, which is described by the Lagrangian ${\cal L}^0_{\rm ALP} + {\cal L}_{\rm HEW}$. Here 
\begin{equation}
\label{t1}
{\cal L}^0_{\rm ALP} = \frac{1}{2} \, \partial^{\mu} a \, \partial_{\mu} a - \frac{1}{2} \, m^2 \, a^2 + \frac{1}{M} \, {\bf E} \cdot {\bf B} \, a
\end{equation}
is the usual ALP lagrangian, where $m$ is the ALP mass and $M$ is a constant with the dimension of an energy, while 
\begin{equation}
\label{t1q}
{\cal L}_{\rm HEW} = \frac{2 \alpha^2}{45 m_e^4} \, \left[ \left({\bf E}^2 - {\bf B}^2 \right)^2 + 7 \left({\bf E} \cdot {\bf B} \right)^2 \right]
\end{equation}
is the Heisenberg-Euler-Weisskopf (HEW) effective lagrangian accounting for the photon one-loop vacuum polarization in the presence of an external magnetic field ($\alpha$ is the fine-structure constant and $m_e$ is the electron mass)~\cite{HEW}. Often 
${\cal L}_{\rm HEW}$ is discarded, but we will see that the inclusion of ${\cal L}_{\rm HEW}$ is essential in our analysis. The other difference is that -- at variance with the case of the axion -- the parameters $m$ and $M$ are supposed to be {\it uncorrelated}, and it is merely supposed that $m < 1 \, {\rm eV}$ and $M \gg G_F^{- 1/2}$ with $G_F^{- 1/2} \simeq 250 \, {\rm GeV}$ denoting the Fermi scale. In this Section we use natural units with $c=\hslash =1$.

Astrophysics provides the strongest bounds on the parameters entering ${\cal L}^0_{\rm ALP}$. The CAST experiment at CERN~\cite{arik2009} sets the quite robust bound $M > 1.14 \cdot 10^{10} \, {\rm GeV}$ for $m < 0.02 \, {\rm eV}$ in agreement with theoretical bounds from stellar evolution~\cite{raffeltbook}, while from the absence of $\gamma$ rays coming from SN 1987A the stronger bound $M >  10^{11} \, {\rm GeV}$ has been derived for $m < 10^{- 10} \, {\rm eV}$ which is however affected by a large uncertainty~\cite{brockwayetal1996}. Actually, the latter bound has been strongly criticized by other authors~\cite{prada2009}, who pointed out additional substantial uncertainties due to the fact that both the flux of ALPs produced in the supernova explosion and their reconversion into photons can vary by large factors, thereby implying that a small violation of such a bound should not be regarded as a flaw of a considered model. Finally, very light ALPs -- which we will see to be our case -- cannot be the galactic dark matter~\cite{massotoldra}, and so the results of the ADMX experiment~\cite{admx} are presently irrelevant.

We consider throughout this paper a monochromatic photon beam of VHE $E$ propagating along the $y$ direction from the centre of the source to us in a generic ionized and magnetized medium where photons can be absorbed but ALPs can not. Below, we describe the mechanism whereby photon-ALP oscillations occur in the beam (a more detailed account can be found in~\cite{dgr}).

Manifestly the last term in Eq. (\ref{t1}) gives rise to the $a \gamma \gamma$ coupling mentioned above. Note that because of the specific form of this coupling $\propto {\bf E} \cdot {\bf B}/M$ only the component of ${\bf B}$ in the $x-z$ plane ${\bf B}_T$ couples to $a$, and in addition all physical results depend only on the combination $B_T /M$ to the extent that ${\cal L}_{\rm HEW}$ can be neglected. Clearly here ${\bf B}$ is the external magnetic field while ${\bf E}$ is the electric field of a propagating photon. 

The first point that deserves concern is that in the approximation $E \gg m$ which is  obviously valid here the beam propagation equation becomes~\cite{raffeltstodolsky1988}
\begin{equation}
\label{t2} 
\left(i \, \frac{d}{d y} + E +  {\cal M} (y,E) \right)  \left(\begin{array}{c}A_x (y) \\ A_z (y) \\ a (y) \end{array}\right) = 0~,
\end{equation}
where $A_x (y)$ and $A_z (y)$ are the two photon linear polarization amplitudes along the $x$ and $z$ axis, respectively, $a (y)$ denotes the ALP amplitude  and ${\cal M} (y,E)$ represents the photon-ALP mixing matrix. 

A very remarkable fact is that Eq. (\ref{t2}) is a Schr\"odinger-like equation with $y$ playing the role of time, provided that the beam is in a pure polarization state. So the relativistic beam in question can formally be treated as a three-level non-relativistic unstable quantum system with in general a not self-adjoint non-stationary Hamiltonian $H (y) = - [E + {\cal M} (y,E) ]$. This circumstance allows us to describe the behaviour of the beam by non-relativistic quantum mechanics. Moreover, we denote by ${\cal U} (y,y_0;E)$ the transfer matrix, namely the solution of Eq. (\ref{t2}) with initial condition ${\cal U} (y_0,y_0;E) = 1$. 

\subsection{Oscillations in an homogeneous magnetic field}

It is convenient to start by supposing that ${\bf B}$ is homogeneous, so that we have the freedom to choose the $z$ axis 
along ${\bf B}_T$. Correspondingly the explicit form of the mixing matrix is~\cite{raffeltstodolsky1988,adler1971}
\begin{equation}
\label{t3}               
{\cal M}_0 (E)  = \left(
\begin{array}{ccc}
\Delta_{\bot} (E) & 0 & 0 \\
0 & \Delta_{\parallel} (E) & \Delta_{a \gamma} \\
0 & \Delta_{ a \gamma} & \Delta_{a a} (E) \\
\end{array}
\right)~,
\end{equation}
since the Faraday effect that would mix $A_x$ and $A_z$ is totally irrelevant at the energies considered in this paper. We denote by $ \lambda_{\gamma} (E)$ the photon mean free path and by $\omega_{\rm pl}$ the plasma frequency, which is related to the electron number density $n_e$ by 
\begin{equation}
\label{t230w} 
\omega_{\rm pl} \simeq 3.69 \cdot 10^{- 11} \left(\frac{n_e}{{\rm cm}^{- 3}} \right)^{1/2} \, {\rm eV}~.
\end{equation}
Then the various $\Delta$ terms entering ${\cal M}_0 (E)$ are 
\begin{equation}
\label{t2qq1} 
\Delta_{\bot} (E) =  - \frac{\omega^2_{\rm pl}}{2 E} + \frac{2 \alpha E}{45 \pi} \left(\frac{B_T}{B_{{\rm cr}}} \right)^2 + \frac{i}{2 \, \lambda_{\gamma} (E)}~,
\end{equation}
\begin{equation}
\label{t2qq2} 
\Delta_{\parallel} (E) = - \frac{\omega^2_{\rm pl}}{2 E} + \frac{3.5 \alpha E}{45 \pi} \left(\frac{B_T}{B_{{\rm cr}}} \right)^2 +  \frac{i}{2 \, \lambda_{\gamma} (E)}~,    
\end{equation}
\begin{equation}
\label{t2qq4} 
\Delta_{a \gamma} = \frac{B_T}{2 M}~,
\end{equation}
\begin{equation}
\label{t2qq5} 
\Delta_{a a} (E) = - \frac{m^2}{2 E}~,
\end{equation}
where $B_{{\rm cr}} \simeq 4.41 \cdot 10^{13} \, {\rm G}$ is the critical magnetic field. The eigenvalues of ${\cal M}_0 (E)$ are
\begin{equation}
\label{gg28a}
\lambda_1(E)=\Delta_{\bot}(E)~,
\end{equation}
\begin{equation}
\label{gg28b}
\lambda_2(E)=\frac{1}{2} \left\{ \Delta_{\parallel}(E) + \Delta_{aa}(E) - \left[ \Bigl(\Delta_{\parallel}(E) - \Delta_{aa}(E) \Bigr)^2 + 4 \Delta^2_{a \gamma} \right]^{1/2} \right\}~,
\end{equation}
\begin{equation}
\label{gg28c}
\lambda_3(E)=\frac{1}{2}\left\{ \Delta_{\parallel}(E) + \Delta_{aa}(E) + \left[ \Bigl(\Delta_{\parallel}(E)-\Delta_{aa}(E) \Bigr)^2 + 4 \Delta^2_{a \gamma} \right]^{1/2} \right\}~,
\end{equation}
and by using the general relations reported in Appendix A of~\cite{dgr} it is straightforward to find the explicit form of the transfer matrix in this case
\begin{equation}
\label{gg28d}
{\cal U}_0 (y,y_0;E) =  e^{i \lambda_1(E) (y - y_0)} \, T_1(E,0) + e^{i \lambda_2(E) (y - y_0)} \, T_2(E,0) + e^{i \lambda_3(E) (y - y_0)} \, T_3(E,0)~,
\end{equation}
where we have set
\begin{equation}
\label{gg28e}
T_1(E,0) \equiv \left(
\begin{array}{ccc}
1 & 0 & 0\\
0 & 0 & 0\\
0 & 0 & 0\\
\end{array}
\right)~,
\end{equation}
\begin{equation}
\label{gg28f}
T_2(E,0) \equiv \left(
\begin{array}{ccc}
0 & 0 & 0\\
0 & \frac{\lambda_3(E)-\Delta_{\parallel}(E)}{\lambda_3(E)-\lambda_2(E)} & - \frac{\Delta_{a \gamma}}{\lambda_3(E)-\lambda_2(E)}\\
0 & - \frac{\Delta_{a \gamma}}{\lambda_3(E)-\lambda_2(E)} & - \frac{\lambda_2(E)-\Delta_{\parallel}(E)}{\lambda_3(E)-\lambda_2(E)}\\
\end{array}
\right)~,
\end{equation}
\begin{equation}
\label{gg28g}
T_3(E,0) \equiv \left(
\begin{array}{ccc}
0 & 0 & 0\\
0 & - \frac{\lambda_2(E)-\Delta_{\parallel}(E)}{\lambda_3(E)-\lambda_2(E)} & \frac{\Delta_{a \gamma}}{\lambda_3(E)-\lambda_2(E)}\\
0 & \frac{\Delta_{a \gamma}}{\lambda_3(E)-\lambda_2(E)} & \frac{\lambda_3(E)-\Delta_{\parallel}(E)}{\lambda_3(E)-\lambda_2(E)}\\
\end{array}
\right)~.
\end{equation}

Consider now the even more particular case in which no absorption is present. Accordingly the photon-ALP conversion 
probability over the distance $y - y_0$ can be computed exactly and reads
\begin{equation}
\label{t8}
P_{0, {\gamma} \to a} (y,y_0;E) = \left(\frac{B_T}{M \, {\Delta}_{\rm osc} (E)}  \right)^2 \,  {\rm sin}^2 \left( \frac{\Delta_{\rm osc} (E) \, (y - y_0)}{2} \right)~,
\end{equation}
where the oscillation wavenumber ${\Delta}_{\rm osc} (E)$ is given by
\begin{equation}
\label{t9}
{\Delta}_{\rm osc} (E) \equiv \left\{\left[\frac{m^2 - {\omega}_{\rm pl}^2}{2  E} + \frac{3.5 \alpha}{45 \pi} \left(\frac{B_T}{B_{\rm cr}} \right)^2 E \right]^2 + \left(\frac{B_T}{M} \right)^2 \right\}^{1/2}~.
\end{equation}
A look at Eqs. (\ref{t8}) and (\ref{t9}) shows that the {\it strong-mixing regime} -- in which the photon-ALP conversion probability becomes maximal and energy-independent -- takes place when the $B_T /M$ term in Eq. (\ref{t9}) dominates. This circumstance can be put into a more explicit form by introducing a low-energy cut-off
\begin{equation}
\label{t9n}
E_L \equiv \frac{M |m^2 - \omega^2_{\rm pl}|}{2 B_T}
\end{equation}
along with a high-energy cut-off
\begin{equation}
\label{t9nn}
E_H \equiv \frac{45 \pi}{3.5 \alpha} \left(\frac{B_{\rm cr}}{B_T} \right)^2 \left(\frac{B_T}{M} \right)~.
\end{equation}
Then the strong-mixing regime occurs in the energy range $E_L < E < E_H$, where {\it the plasma contribution, the ALP mass term and the QED one-loop effect are negligible} and should be discarded from the mixing matrix. The strong-mixing regime does not exist for $E_H < E_L$.

Our choice to take the $z$ axis along ${\bf B}_T$ has the advantage to make the above equations simpler and more eloquent, but in view of our later analysis it is essential to contemplate the situation in which ${\bf B}_T$ forms a non vanishing angle $\psi$ with the 
$z$ axis. So, we proceed to work out the explicit form of the previous equation in the general case. Accordingly, the mixing matrix arises from ${\cal M}_0 (E)$ through the similarity transformation
\begin{equation}
\label{t5L28q}
{\cal M} (E,\psi) = V^{\dagger} (\psi) \, {\cal M}_0 (E) \, V (\psi)~, 
\end{equation}
effected by the rotation matrix in the $x - z$ plane
\begin{equation}
\label{gg28q1}
V(\psi) =\left(
\begin{array}{ccc}
{\rm cos} \, \psi & -{\rm sin} \, \psi & 0\\
{\rm sin} \, \psi  & {\rm cos} \, \psi & 0\\
0 & 0 & 1\\
\end{array}
\right)~,
\end{equation}
so that we get~\cite{gf}
\begin{equation}
\label{t3G}
{\cal M} (E,\psi) \equiv \left(
\begin{array}{ccc}
\Delta_{xx} (E,\psi) & \Delta_{xz} (E,\psi) & \Delta_{a \gamma} \, {\rm sin} \, \psi \\
\Delta_{zx} (E,\psi)& \Delta_{zz} (E,\psi) & \Delta_{a \gamma} \, {\rm cos} \, \psi \\
\Delta_{a \gamma} \, {\rm sin}  \, \psi & \Delta_{ a \gamma} \, {\rm cos} \, \psi & \Delta_{a a} (E) \\
\end{array}
\right)~,
\end{equation}
with
\begin{equation}
\label{t9nn1}
\Delta_{xx} (E,\psi) \equiv \Delta_{\parallel} (E) \, {\rm sin}^2 \, \psi + \Delta_{\bot} (E) \, {\rm cos}^2 \, \psi~,
\end{equation}
\begin{equation}
\label{t9nn2}
\Delta_{xz} (E,\psi) = \Delta_{zx} (E) \equiv \left(\Delta_{\parallel} (E) - \Delta_{\bot} (E) \right) {\rm sin} \, \psi \, {\rm cos} \, \psi~,
\end{equation}
\begin{equation}
\label{t9nn3}
\Delta_{zz} (E,\psi) \equiv \Delta_{\parallel} (E) \, {\rm cos}^2 \, \psi + \Delta_{\bot} (E) \, {\rm sin}^2 \, \psi~.
\end{equation}
Correspondingly, the transfer matrix becomes
\begin{equation}
\label{gg28dd}
{\cal U} (y,y_0;E;\psi) =  V^{\dagger} (\psi) \, {\cal U}_0 (y,y_0;E) \, V (\psi)~.
\end{equation}
Such an expression is fairly complicated in general, and we shall work it out whenever necessary.

So far, we have supposed that the photons in the initial beam have a definite polarization. However, in the situation under consideration such a polarization is unknown to a considerably extent, and hence it is safe to regard  the beam initially as unpolarized. Therefore, it must be described by the (generalized) density matrix $\rho (y)$ which -- thanks to the above analogy with non-relativistic quantum mechanics -- obeys the Von Neumann-like equation
\begin{equation}
\label{t5}
i \frac{d \rho (y)}{d y} = \rho (y) \, {\cal M}^{\dagger} (y,E) - {\cal M} (y,E) \, \rho (y)
\end{equation}
associated with Eq. (\ref{t2}). Then it follows that even if $ {\cal M}^{\dagger} (y,E) \neq {\cal M} (y,E)$ the solution of Eq. (\ref{t5}) is given by 
\begin{equation}
\label{t5L}
\rho (y,E) = {\cal U} (y, y_0;E) \, \rho_{\rm in} \, {\cal U}^{\dagger}(y, y_0;E)~, 
\end{equation}
where $\rho_{\rm in} \equiv \rho (y_0)$ is the initial beam state. Accordingly, the probability that the beam will be found in the final state $\rho_{\rm fin}$ at $y$ is given by
\begin{equation}
\label{t72805}
P_{\rho_{\rm in} \to \rho_{\rm fin}} (y,y_0;E) = {\rm Tr} \Bigl( \rho_{\rm fin} \, {\cal U} (y,y_0;E) \, \rho_{\rm in} \, {\cal U}^{\dagger}(y, y_0;E) \Bigr)~,
\end{equation}
where it is assumed that ${\rm Tr} \, \rho_{\rm in} = {\rm Tr} \, \rho_{\rm fin} =1$.

Clearly in the particular case treated here Eqs. (\ref{t5}), (\ref{t5L}) and (\ref{t72805}) hold for ${\cal M} (y,E) \to {\cal M} (\psi,E)$ and ${\cal U} (y, y_0;E) \to {\cal U} (y,y_0;E;\psi)$.

\subsection{Oscillations in an inhomogeneous magnetic field}

As we will see, in the present model of PKS 1222+216 the beam crosses four regions with very different properties before being detected, and only in the first one ${\bf B}$ can be taken as homogeneous in first approximation. In the second region, ${\bf B}$ has a smooth $y$-dependence and in this case the beam propagation equation can be solved exactly owing to a drastic simplification of the $\Delta$ terms entering the mixing matrix (we find it more natural to address this issue in Section III-C where the properties of the region in question are discussed, which in turn dictate the form of the $\Delta$ terms). In the third region ${\bf B}$ possesses a turbulent structure which is currently modeled as random domain-like network, and this may or may not be the case in the fourth region (more about this, later). 

Specifically, in the simplest situation all domains have the same size set by the coherence length of the magnetic field and the strength of ${\bf B}$ is the same in every domain, but the orientation of ${\bf B}$ changes randomly from one domain to the next (however a more general situation has to be considered in the cosmological context, but these complications will be discussed later).  Therefore, inside the $n$-th generic domain ($1 \leq n \leq N$) the mixing matrix in Eq. (\ref{t2}) takes the form
\begin{equation}
\label{t3G}
{\cal M}_n (E,\psi_n) \equiv \left(
\begin{array}{ccc}
\Delta_{xx} (E,\psi_n) & \Delta_{xz} (E,\psi_n) & \Delta_{a \gamma} \, {\rm sin} \, \psi_n \\
\Delta_{zx} (E,\psi_n)& \Delta_{zz} (E,\psi_n) & \Delta_{a \gamma} \, {\rm cos} \, \psi_n \\
\Delta_{a \gamma} \, {\rm sin}  \, \psi_n & \Delta_{ a \gamma} \, {\rm cos} \, \psi_n & \Delta_{a a} (E) \\
\end{array}
\right)~,
\end{equation}
where the various $\Delta (E,\psi_n)$ quantities are just given by Eqs. (\ref{t9nn1}), (\ref{t9nn2}) and (\ref{t9nn3}). Manifestly 
$\psi_n$ is the angle between ${\bf B}_T^{(n)}$ and the a fiducial $z$ axis taken to be the same for all domains. Denoting by ${\cal U}_n \left(E;\psi_n \right)$ the transfer matrix in the $n$-th domain -- which is derived through the same steps as in the previous case --  its explicit form just follows from Eq. (\ref{gg28dd}) with $y - y_0$ denoting the domain size.

Moreover, according to quantum mechanics the transfer matrix for the whole network of $N$ domains is 
\begin{equation}
\label{t9nK26a}
{\cal U}_{\rm random} \left(E; \psi_1, ... , \psi_N \right) = \prod^{N}_{n = 1} \, {\cal U}_n \left(E; \psi_n \right)~.
\end{equation}
Further, it proves convenient to denote by ${\cal U}_{\rm smooth} (E)$ the transfer matrix corresponding to the first two regions. Hence, the overall beam propagation is described by ${\cal U}_{\rm random} \left(E; \psi_1, ... , \psi_N \right) \, {\cal U}_{\rm smooth} (E)$. As a consequence, the probability that the beam emitted in the state $\rho_{\rm in}$ at $y_0$ will be detected in the state $\rho_{\rm fin}$ at $y$ for {\it fixed} orientations $\psi_1, ... , \psi_N$ of ${\bf B}$ in every domain is 
\begin{eqnarray}
\label{t7280527bLl}
P_{\rho_{\rm in} \to \rho_{\rm fin}} (y,y_0;E;\psi_1, ... , \psi_N) =  \ \ \ \ \ \ \ \ \ \ \ \ \ \ \ \ \ \ \ \ \ \ \ \ \ \ \ \   \\  \nonumber  
= {\rm Tr} \Bigl( \rho_{\rm fin} \, {\cal U}_{\rm random} \left(E; \psi_1, ... , \psi_N \right) \, {\cal U}_{\rm smooth} (E)  \, \rho_{\rm in} \, {\cal U}^{\dagger}_{\rm smooth} (E) \, {\cal U}^{\dagger}_{\rm random} \left(E; \psi_1, ... , \psi_N \right) \Bigr)~,
\end{eqnarray}
where it is assumed again that ${\rm Tr} \, \rho_{\rm in} = {\rm Tr} \, \rho_{\rm fin} =1$.

Now, owing to the turbulent nature of the magnetic field in question the angles $\psi_n$ ($1 \leq n \leq N$) are to be regarded as independent random variables in the range $1 \leq \psi_n \leq 2 \pi$. As a consequence, the physical detection probability for the considered beam arises by averaging the last equation over all angles, namely we have
\begin{equation}
\label{t7280527b}
P_{\rho_{\rm in} \to \rho_{\rm fin}} (y,y_0;E) = \Bigl\langle P_{\rho_{\rm in} \to \rho_{\rm fin}} (y,y_0;E; \psi_1, ... , \psi_N) \Bigr\rangle_{\psi_1, ... , \psi_N}~.
\end{equation}

More specifically, since the photon polarization cannot be measured at the considered energies, we have to sum Eq. (\ref{t7280527b}) over the two final polarization states  
\begin{equation}
\label{a9s11A}
{\rho}_x = \left(
\begin{array}{ccc}
1 & 0 & 0 \\
0 & 0 & 0 \\
0 & 0 & 0 \\
\end{array}
\right)~,
\end{equation}
\begin{equation}
\label{a9s11B}
{\rho}_z = \left(
\begin{array}{ccc}
0 & 0 & 0 \\
0 & 1 & 0 \\
0 & 0 & 0 \\
\end{array}
\right)~.
\end{equation}
In addition, we suppose for simplicity that the emitted beam consists $100 \, \%$ of unpolarized photons, so that the initial beam state is described by
\begin{equation}
\label{a9s11C}
{\rho}_{\rm unpol} = \frac{1}{2}\left(
\begin{array}{ccc}
1 & 0 & 0 \\
0 & 1 & 0 \\
0 & 0 & 0 \\
\end{array}
\right)~.
\end{equation}
Therefore Eq. (\ref{t7280527b}) ultimately takes the form
\begin{eqnarray}
\label{k3lwf1w1WW}
P_{\gamma \to \gamma} (y,y_0;E)  =  \Big\langle P_{\rho_{\rm unpol} \to \rho_x} \left(y,y_0;E; \psi_1, ... , \psi_{N} \right) \Big\rangle_{\psi_1, ... , \psi_{N}} +   \\  \nonumber
+ \, \Big\langle P_{\rho_{\rm unpol} \to \rho_z} \left(y, y_0;E; \psi_1, ... , \psi_{N} \right) 
\Big\rangle_{\psi_1, ... , \psi_{N}}~,   \ \ \ \ \ \ \ \ \ \ \ 
\end{eqnarray}
which gives the photon survival probability.

The actual evaluation of $P_{\gamma \to \gamma} \left(E \right)$ goes as follows. In the first place we have to know ${\cal U}_{\rm smooth} (E)$. Next, we arbitrarily choose $\psi_n$ in each domain and so we can evaluate ${\cal U}_n \left(\psi_n,E \right)$ for a given energy $E$. Thanks to Eq. (\ref{t9nK26a}) and the next one, we find the photon survival probability for a single realization of the propagation process. We repeat these steps $5000$ times, by randomly varying all angles $\psi_n$ each time, thereby generating $5000$ random realizations of the propagation process. Finally, we average the resulting photon survival probabilities over all these realizations of the propagation process, thereby accomplishing the average process in the last equation. We find in this way the physical photon survival probability $P^{\rm ALP}_{\gamma \to \gamma} \left(E \right)$ when ALP effects are included.

\section{A model for PKS 1222+216}

We now proceed to build up our model for PKS 1222+216. We start by explicitly showing why it should not be observed by MAGIC according to conventional physics. We next proceed to discuss photon-ALP conversions in the beam propagation from the central region of PKS 1222+216 to the Earth. Basically, four different regions crossed by the beam are identified, and in each of them we evaluate the relevant transfer matrix. All these pieces of information will be put together in Section IV in order to find out the resulting physical effect.

We stress that in this investigation we will make some rather rough assumptions even because of the lack of knowledge of a few properties of the source (we defer a more detailed modelling to a future publication).

\subsection{Observations and setup}

In the first place we need to know the relevant physical parameters. We assume a disk luminosity $L_{\rm D} \simeq 1.5\cdot 10^{46}$ erg s$^{-1}$, a radius of the BLR $R_{\rm BLR} \simeq 0.23 \, {\rm pc}$, and standard values for cloud number density $n_c \simeq10^{10} \, {\rm cm}^{-3}$ and temperature $T_c \simeq 10^4 \, {\rm K}$ of the BLR (see e.g.~\cite{tavecchiomazin2009}). 
The adopted disk luminosity (a factor of $\sim3$ less than that derived in \cite{tavecchioetal2011} by the observed optical-UV continuum interpreted as the direct emission from the disk) is calculated -- following the method outlined in e.g.~\cite{cpg97} -- from the luminosity of the broad emission lines (H$\alpha$, H$\beta$, MgII) recently obtained in~\cite{farina} using several optical spectra taken in the period 2008-2011.  The adopted value of $R_{\rm BLR}$ is derived from the measured line width and the black hole mass again in~\cite{farina}.

Since the filling factor of the clouds is small, the average electron number density $n_e$ relevant for the beam propagation is much smaller than $n_c$. Models attributing the confinement of the clouds to a hot $T_e\simeq 10^{7-8}$ K external medium in pressure equilibrium with the clouds yield $n_e \simeq 10^{6-7} \, {\rm cm}^{- 3}$ (see e.g.~\cite{kroliketal1981}). However, the presence of such a hot confining medium is disfavored by the lack of the necessarily expected bright X-ray emission. So, an extra contribution to the pressure confining the BLR clouds is expected and the most likely one is due to a magnetic field with strength $B \sim 1 \, {\rm G}$~\cite{rees1987} (for a review, see~\cite{netzer2008}). As a consequence, $n_e$ gets considerably reduced, perhaps to values as low as $n_e \simeq 10^4 \, {\rm cm}^{- 3}$. 

In order to bring out quite explicitly the problem posed by the VHE observation of PKS 1222+216, we compute the optical depth 
$\tau(E)$ of the beam photons in the BLR according to conventional physics, which amounts to consider only the process 
$\gamma \gamma \to e^+ e^-$. We follow the same procedure developed in~\cite{tavecchiomazin2009}, to which the reader is referred for a full description. The optical depth is given by
\begin{equation}
\tau(E)=\int d\Omega \int d\epsilon \int dx ~ n_{\rm ph}(\epsilon,\Omega,x) \, \sigma _{\gamma\gamma}(E, \epsilon, \mu) (1-\mu )~, 
\label{tau}
\end{equation}
\noindent
where $E$ is the energy of a $\gamma$ ray, $x$ is the distance from the center of the BLR, $\mu \equiv \cos \theta$ where $\theta$ is the scattering angle between the $\gamma$ ray and the soft photon of energy $\epsilon$, $d\Omega = - 2 \pi d\mu$ while $n_{\rm ph}(\epsilon,\Omega,x)$ is the spectral number density of the BLR radiation field per unit solid angle at position $x$ and the $\gamma \gamma$ pair-production cross-section reads~\cite{pairproduction1}
\begin{equation} 
\label{eq.sez.urto}
\sigma_{\gamma \gamma}(E,\epsilon,\mu)  \simeq 1.25 \cdot 10^{-25} \left(1-\beta^2 \right) \left[2 \beta \left( \beta^2 -2 \right) 
+ \left( 3 - \beta^4 \right) \, {\rm ln} \left( \frac{1+\beta}{1-\beta} \right) \right] {\rm cm}^2~,
\end{equation}
which depends on $E$, $\epsilon$ and $\mu$ only through the dimensionless parameter
\begin{equation} 
\label{eq.sez.urto01012011q}
\beta(E,\epsilon,\mu) \equiv \left[ 1 - \frac{2 \, m_e^2 \, c^4}{E \epsilon \left(1-\mu \right)} \right]^{1/2}~. 
\end{equation}
Regarding $E$ as an independent variable, the process is kinematically allowed for
\begin{equation} 
\label{eq.sez.urto01012011}
\epsilon > {\epsilon}_{\rm thr}(E,\mu) \equiv \frac{2 \, m_e^2 \, c^4}{ E \left(1-\mu \right)}~.
\end{equation}
The quantity $n_{\rm ph}(\epsilon,\Omega,x)$ concerning the target soft photons emitted by the photo-ionized BLR clouds is
calculated using the standard photo-ionization code CLOUDY as in \cite{tavecchio2008} using the input parameters listed above. 

\begin{figure}[h]
\centering
\vspace{- 2truecm}
\includegraphics[width=.55\textwidth]{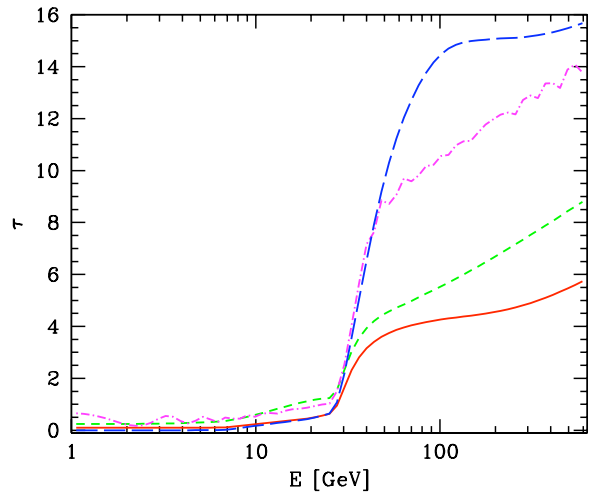}
\vspace{-3 truecm}
\caption{\label{fig:surf} Effective optical depth plotted as a function of rest-frame energy for VHE photons propagating in the BLR of PKS 1222+216. The blue long-dashed line corresponds to the process $\gamma \gamma \to e^+ e^-$. The other three lines pertain to our model containing ALPs. Specifically, the violet dashed-dotted line corresponds to $(B = 2 \, {\rm G}, M = 4 \cdot 10^{11} \, {\rm GeV})$, the green short-dashed line to $(B = 0.4 \, {\rm G}, M = 1.5 \cdot 10^{11} \, {\rm GeV})$ and the red solid line to $(B = 0.2 \, {\rm G}, M = 7 \cdot 10^{10} \, {\rm GeV})$ (more about this, later and in Section IV-A).}
\end{figure}

The result is plotted as the blue long-dashed line in Fig. \ref{fig:surf}, which shows that PKS 1222+216 should indeed be totally invisible in the energy range $70 - 400 \, {\rm GeV}$ where it has instead been detected by MAGIC.

The derived $\tau(E)$ is affected by some degree of uncertainty, which is a direct consequence of  the uncertainty associated with some of the input parameters, in particular the luminosity of the disk and the radius of the BLR. For instance, a disk luminosity smaller by a factor of $\sim \, 3$ was derived in~\cite{fermi} based on the luminosity of the broad H$\beta$ line measured from an old optical spectrum. Moreover, an uncertainty in $L_{\rm D}$ also affects the value of the radius of the BLR, since  it is customarily estimated by assuming the empirical relation $R_{\rm BLR}\propto L_{\rm D}^{1/2}$ (see e.g.~\cite{canonical} and references therein). We can roughly summarize the effect of the error associated with $L_{\rm D}$ and $R_{\rm BLR}$ on the optical depth by the following chain involving only the relevant quantities: $\tau \propto R_{\rm BLR}\, n_{\rm ph} \propto R_{\rm BLR}\,  L_{\rm BLR}/R_{\rm BLR}^2 \propto L_{\rm D}^{1/2}$, which indeed leads to $\tau \propto L_{\rm D}^{1/2}$. Therefore, the final impact of these uncertainties is moderate. In fact, assuming   $L_{\rm D} \simeq  5 \cdot 10^{45}$ erg s$^{-1}$ as in~\cite{fermi}, an optical depth comparable to our result is obtained in~\cite{dermer}. 

A possible complication is the presence within the BLR of disk photons scattered by the high-temperature gas assumed to fill the region between the clouds. The total scattering optical depth associated with such a gas with the density $n_e \simeq 10^4 \, {\rm cm}^{- 3}$ assumed above is $\tau_{\rm sc}=\sigma _{\rm T} R_{\rm BLR} n_e\simeq 2.5\cdot 10^{-3}$. A detailed calculation of the optical depth of $\gamma$ rays associated with the scattered disk photons has been made in~\cite{dermer}, where it is found that  the maximum absorption caused by this component is localized at $\sim 200$ GeV, with a corresponding optical depth $\tau \sim 70 \, \tau_{\rm sc}$. Clearly $\tau_{\rm sc}=2.5 \cdot 10^{-3}$ entails $\tau \simeq 0.2$, thus showing that this contribution to the total $\tau(E)$ can be safely neglected.

As pointed out in Section I, our proposal to avoid the problem of the huge optical depth is framed within the standard blazar model for photon production, but in addition we assume the existence of ALPs $a$ with parameters allowing for efficient conversions $\gamma \to a$ and $a \to  \gamma$. Basically, we envisage that a large fraction of VHE photons produced as usual near the central engine become ALPs before reaching the BLR, thereby crossing it totally unimpeded. Outside the BLR most of the ALPs are supposed to become VHE photons again in the large scale jet and in the host galaxy. Because the beam traverses the extragalactic space, the possibility of photon-ALP oscillations in this region has also to be taken into account. Finally, we recall that the external magnetic field is the crucial quantity that triggers photon-ALP conversions.

As a preliminary step, we find it very useful to rewrite Eqs. (\ref{t9n}) and (\ref{t9nn}) in the following more convenient form 
\begin{equation}
\label{t9nK}
E_L \simeq 25 \left| \left(\frac{m}{10^{- 10} \, {\rm eV}} \right)^2 - 0.13 \left(\frac{n_e}{{\rm cm}^{- 3}} \right) \right| \left(\frac{G}{B_T} \right) \left(\frac{M}{10^{11} \, {\rm GeV}} \right) {\rm eV}  
\end{equation}
and
\begin{equation}
\label{t9nnK}
E_H \simeq 2.1 \left(\frac{G}{B_T} \right) \left(\frac{10^{11} \, {\rm GeV}}{M} \right) {\rm GeV}~, 
\end{equation}
so that we can very easily find out when the strong-mixing regime occurs. Moreover, we observe that in the mixing matrix the plasma contribution is negligible with respect to the QED one-loop contribution for
\begin{equation}
\label{t9nnK1}
n_e \ll 7.8 \cdot 10^{11} \left(\frac{E}{100 \, {\rm GeV}} \right)^2 \left(\frac{B_T}{{\rm G}} \right)^2 {\rm cm}^{- 3}~, 
\end{equation}
thanks to Eqs. (\ref{t230w}) and (\ref{t2qq1}). Of course, condition (\ref{t9nnK1}) is relevant outside the strong-mixing regime.

Below, we will address photon-ALP conversions in the various regions crossed by the beam.

\subsection{Photon-ALP oscillations before the BLR}

We start by considering the inner part of the blazar, namely the region extending from the centre to $R_{\rm BLR} \simeq 0.23 \, {\rm pc}$, to be referred to as region 1. 

The magnetic field profile along the jet is well known to decrease outwards but unfortunately the presence of strong shocks and relativistic winds makes a precise estimate of the strength profile of ${\bf B}$ for distances smaller than $R_{\rm BLR}$ practically impossible. An analysis based on a highly idealized description relying on a one-zone, homogeneous leptonic model with external photons yields for the strength of ${\bf B}$ at the base of the jet $B \simeq 2.2 \, {\rm G}$~\cite{celotti2008}. Therefore, we feel it realistic in this investigation to assume its strength to be constant and equal to its average value from the centre to the BLR, which we take for definiteness $B \simeq 0.2 \, {\rm G}$. Moreover, owing to the complicated morphology of ${\bf B}$, we similarly suppose that its average direction from the centre to the BLR is nonvanishing and we assume it to be the same everywhere in the considered region (equal to its average direction), so that we are dealing with a {\it homogeneous} ${\bf B}$ having an unknown direction. Because the photon-ALP conversions vanish if ${\bf B}$ is just along the beam while it is maximal for ${\bf B}$ transverse to the beam, we suppose in line with our heuristic attitude that ${\bf B}$ is on average at an angle of $45^{\circ}$ with the beam direction. Therefore we have $B_T \simeq 0.14 \, {\rm G}$.

A natural question is to find out the energy range $E_L < E < E_H$ in which the strong-mixing regime takes place. As a working hypothesis we assume $m < \omega_{\rm pl}$, so that $E_L$ becomes independent of $m$, and in addition it follows that we are dealing with a {\it very light} ALP just like in previous work~\cite{hooperserpico,dmr,darma,dgr,simetetal2008,prada2009} (more about this, later). Hence from Eq. (\ref{t9nK}) we find $E_L \simeq 0.23 \, {\rm MeV}$ for $n_e \simeq 10^4 \, {\rm cm}^{- 3}$ while $E_L \simeq 0.23 \, {\rm GeV}$ for $n_e \simeq 10^7 \, {\rm cm}^{- 3}$. Further, thanks to Eq. (\ref{t9nnK}) the robust CAST bound entails $E_H < 136 \, {\rm GeV}$ and the controversial SN 1987A bound implies $E_H < 15 \, {\rm GeV}$. Thus, we end up with the conclusion that for {\it Fermi}/LAT observations we are in the strong-mixing regime but for MAGIC observations we are {\it not}. Yet, we will see that the latter fact does not undermine the relevance of photon-ALP oscillations well above $E_H$. Below, we put ourselves in the general case where the strong-mixing regime does not take place in order to have a uniform treatment for both the {\it Fermi}/LAT and MAGIC observations:  Of course, the simplifications characteristic of the strong-mixing regime automatically show up whenever it takes place. 

An additional issue concerns the relevance of the QED one-loop effect. We know that it is negligible in the strong-mixing regime, but otherwise it can be important. Since photon absorption is independent of ${\bf B}$ it can presently be discarded, so that the relevant ${\bf B}$-dependent quantity governing the photon-ALP conversion probability is the oscillation wavenumber ${\Delta}_{\rm osc} (E)$. Then a look at Eq. (\ref{t9}) immediately shows that the QED one-loop is unimportant with respect to the standard magnetic contribution at energies $E < E_H$, namely for 
\begin{equation}
\label{mrds1}
E < 15 \left(\frac{10^{11} \, {\rm GeV}}{M} \right) {\rm GeV}~,
\end{equation}
as expected. So, for $M$ rather close to $10^{11} \, {\rm GeV}$ -- which we will see to be indeed our case -- the QED one-loop effect is negligible for {\it Fermi}/LAT observations ($0.3 - 3\, {\rm GeV}$) but of paramount importance for MAGIC observations ($70 - 400 \, {\rm GeV}$).

Now, owing to Eq. (\ref{t230w}) $n_e \simeq 10^4 \, {\rm cm}^{- 3}$ gives $\omega_{\rm pl} \simeq 3.69 \cdot 10^{- 9} \, {\rm eV}$ whereas $n_e \simeq 10^7 \, {\rm cm}^{- 3}$ yields $\omega_{\rm pl} \simeq 1.17 \cdot 10^{- 7} \, {\rm eV}$, which correspondingly imply $m < 3.69 \cdot 10^{- 9} \, {\rm eV}$ and $m <  1.17 \cdot 10^{- 7} \, {\rm eV}$. Furthermore, for our choice $B_T \simeq 0.14 \, {\rm G}$ condition (\ref{t9nnK1}) implies that the plasma contribution is always negligible with respect to the QED one-loop contribution both for {\it Fermi/LAT} and for MAGIC observations (of course in the former case the plasma contribution is {\it a fortiori} negligible with respect to the standard magnetic contribution). Hence in the mixing matrix the plasma effect can be discarded. Manifestly, within the strong-mixing regime the ALP mass term in the mixing matrix has to be neglected, but it is easy to check that for the above choice of the parameters the same situation is true to leading order even at higher energies where the strong-mixing regime does not occur.

Finally, we have 
\begin{equation}
\label{mrds31}
\lambda_{\gamma} (E) = \frac{R_{\rm BLR}}{\tau(E)}~,
\end{equation}
where $\tau(E)$ is the optical depth for $\gamma \gamma \to e^+ e^-$ plotted by the blue long-dashed line in Fig. \ref{fig:surf} and evaluated as in~\cite{tavecchiomazin2009}. 

Thus, on account of Eqs. (\ref{t2qq1}), (\ref{t2qq2}), (\ref{t2qq4}), (\ref{t2qq5}) and (\ref{mrds31}) we find that to leading order the elements of the mixing matrix (\ref{t3}) are
\begin{equation}
\label{t2qq1k} 
\Delta_{\bot} (E) = \frac{2 \alpha E}{45 \pi} \left(\frac{B_T}{B_{{\rm cr}}} \right)^2 + \frac{i \, \tau(E)}{2 \, R_{\rm BLR}} \simeq 
10^{- 24} \left[\left(\frac{E}{{\rm GeV}} \right) + 13.9 \, i \, \tau(E) \right] {\rm eV}~,
\end{equation}
\begin{equation}
\label{t2qq2k} 
\Delta_{\parallel} (E) = \frac{3.5 \alpha E}{45 \pi} \left(\frac{B_T}{B_{{\rm cr}}} \right)^2 + \frac{i \, \tau(E)}{2 \, R_{\rm BLR}} \simeq 10^{- 24} \left[1.75 \left(\frac{E}{{\rm GeV}} \right) + 13.9 \, i \, \tau(E) \right] {\rm eV}~,    
\end{equation}
\begin{equation}
\label{t2qq4k} 
\Delta_{a \gamma} = \frac{B_T}{2 M} \simeq 1.37 \cdot 10^{- 23} \left(\frac{10^{11} \, {\rm GeV}}{M} \right) {\rm eV}~,
\end{equation}
\begin{equation}
\label{t2qq5k} 
\Delta_{a a} (E) = 0~.
\end{equation}
It is now a matter of simple algebra to evaluate the corresponding eigenvalues $\lambda_1(E)$, $\lambda_2(E)$ and 
$\lambda_3(E)$ as well as the matrices $T_1(E,0)$, $T_2(E,0)$ and $T_3(E,0)$ entering Eq. (\ref{gg28d}), so that we ultimately get the transfer matrix ${\cal U}_1 (R_{\rm BLR},0;E)$.

\subsection{Photon-ALP oscillations in the large scale jet}

Let us next focus our attention on the region surrounding the BLR along the line of sight -- to be referred to as region 2 -- namely on the jet beyond the parsec scale or more precisely beyond $R_{\rm BLR}$. The main question concerns the behavior of ${\bf B}$. At variance with the inner region, shocks and winds are expected to be here less relevant so that we can attempt to figure out the $y$-dependence of ${\bf B}$. As is well known, a poloidal field behaves as $B (y) \propto y^{- 2}$ whereas a toroidal field goes like 
$B (y) \propto y^{- 1}$ \cite{begelman}. Clearly, since we are at a sufficiently large distance from the centre the toroidal field dominates, and so it seems more plausible to assume $B (y) \propto y^{- 1}$ (see also \cite{gabuzda}). As a consequence, in the region 2 we adopt the profile
\begin{equation}
\label{mmr1} 
B_T(y) \simeq 0.14 \, \left(\frac{R_{\rm BLR}}{y} \right) \, {\rm G} \simeq 3.22 \cdot 10^{- 2} \left(\frac{{\rm pc}}{y} \right) {\rm G}~.
\end{equation}
However, we have checked that even by taking $B (y) \propto y^{- 2}$ the corresponding change is small, thereby showing that  the choice of the exact profile of ${\bf B}$ has a minor impact on the final result.

What is the actual size of the region 2? As we shall see, the typical strength of the turbulent magnetic field in the host elliptical galaxy is about $5 \, \mu{\rm G}$, and so it looks natural to define the outer edge $R_*$ of region 2 as the galactocentric distance where $B_T(y)$ in Eq. (\ref{mmr1}) reaches the value of $5 \, \mu{\rm G}$. Accordingly, we get $R_* \simeq 6.7 \, {\rm kpc}$. 

A relevant question is whether the strong-mixing regime takes place throughout the whole energy range -- namely for $0.3 \, {\rm GeV} < E < 400 \, {\rm GeV}$ -- over region 2. For this to be the case, we must have both $E_L < 0.3 \, {\rm GeV}$ and $E_H > 400 \, {\rm GeV}$. Explicitly, by combining Eqs. (\ref{t9nK}) and (\ref{t9nnK}) with Eq. (\ref{mmr1}) we obtain 
\begin{equation}
\label{t9nKQ}
y < 3 \cdot 10^3 \left(\frac{{\rm cm}^{- 3}}{n_e} \right) \left(\frac{10^{11} \, {\rm GeV}}{M} \right) {\rm kpc}
\end{equation}
and 
\begin{equation}
\label{t9nKQ1}
y > 6.1 \left(\frac{M}{10^{11} \, {\rm GeV}} \right) {\rm pc}~.
\end{equation}
So, we see that even by taking the average electron density as large as $n_e \simeq 10^2 \, {\rm cm}^{- 3}$ we reach the conclusion that the strong-mixing regime takes place over more than $99 \, \%$ of region 2 for $M$ rather close to $10^{11} \, {\rm GeV}$. Thus, it will be assumed {\it tout court}. Consequently, the plasma contribution, the ALP mass term and the QED one-loop contribution should be dropped.

Since in region 2 absorption effects are negligible, owing to the above conclusions from Eqs. (\ref{t2qq1}), (\ref{t2qq2}), (\ref{t2qq4}) and (\ref{t2qq5}) we ultimately obtain
\begin{equation}
\label{t2qq1kc} 
\Delta_{\bot} (E) = 0~,
\end{equation}
\begin{equation}
\label{t2qq2kc} 
\Delta_{\parallel} (E) = 0~,    
\end{equation}
\begin{equation}
\label{t2qq4kc} 
\Delta_{a \gamma} = \frac{B_T}{2 M} \simeq 3.1 \cdot 10^{- 24} \left(\frac{{\rm pc}}{y} \right) \left(\frac{10^{11} \, {\rm GeV}}{M} \right) {\rm eV}~,
\end{equation}
\begin{equation}
\label{t2qq5kc} 
\Delta_{a a} (E) = 0~. 
\end{equation}
In this case, the transfer matrix can be directly obtaining by explicitly solving the beam propagation equation and reads 
\begin{equation}
\label{}
 {\cal U}_2 (R_*,R_{\rm BLR};E)= \left(
\begin{array}{ccc}
1 & 0 & 0\\
0 & {\rm cos} \left(  \frac{B_T(R_{\rm BLR}) \, R_{\rm BLR}}{2 M} \, {\rm ln} \frac{R_{*}}{R_{\rm BLR}} \right) & i \, {\rm sin} \left(  \frac{B_T(R_{\rm BLR}) \, R_{\rm BLR}}{2 M} \, {\rm ln}\frac{R_{*}}{R_{\rm BLR}} \right)\\
0 & i \, {\rm sin} \left(  \frac{B_T(R_{\rm BLR}) \, R_{\rm BLR}}{2 M} \, {\rm ln} \frac{R_{*}}{R_{\rm BLR}} \right) & {\rm cos} \left(  \frac{B_T(R_{\rm BLR}) \, R_{\rm BLR}}{2 M} \, {\rm ln}\frac{R_{*}}{R_{\rm BLR}} \right)\\
\end{array}
\right)~,
\end{equation}
where of course $B_T(R_{\rm BLR})$ is supplied by Eq. (\ref{mmr1}).

\subsection{Photon-ALP oscillations in the host galaxy}

FSRQs are hosted by elliptical galaxies, whose ${\bf B}$ is very poorly known. Nevertheless, it has been argued~\cite{mossshukurov} that supernova explosions and stellar motion give rise to a turbulent ${\bf B}$ which can be modeled by a domain-like structure, with average strength $5 \, \mu{\rm G}$ and domain size equal to $150 \, {\rm pc}$. We stress that these two quantities are the same for all domains. Correspondingly, we define region 3 as the spherical section with inner radius $R_* \simeq 6.7 \, {\rm kpc}$ and outer radius $R_{\rm host}$. 

Because absorption is presently irrelevant, the transfer matrix ${\cal U}_3 \left(R_{\rm host}, R_*; E; \phi_1, . . . , \phi_{N_3} \right)$ just follows from the discussion in Section II-B with $\psi_n \to \phi_n$ and $N \to N_3$. Furthermore, it is trivial to check that in region 3 the strong-mixing regime takes place and so the plasma contribution, the ALP mass term and the QED one-loop contribution 
are totally irrelevant. Hence, from Eqs. (\ref{t2qq1}), (\ref{t2qq2}), (\ref{t2qq4}) and (\ref{t2qq5}) in every domain we have
\begin{equation}
\label{t2qq1kcd} 
\Delta_{\bot} (E) = 0~,
\end{equation}
\begin{equation}
\label{t2qq2kcd} 
\Delta_{\parallel} (E) = 0~,    
\end{equation}
\begin{equation}
\label{t2qq4kcd} 
\Delta_{a \gamma} = \frac{B_T}{2 M} \simeq 3.4 \cdot 10^{- 28} \left(\frac{10^{11} \, {\rm GeV}}{M} \right) {\rm eV}~,
\end{equation}
\begin{equation}
\label{t2qq5kcd} 
\Delta_{a a} (E) = 0~, 
\end{equation}
and so in the $n$-th generic domain ($1 \leq n \leq N_3$) the mixing matrix (\ref{t3G}) takes the explicit form
\begin{equation}
\label{t3GN}
{\cal M}_n (E,\phi_n) = \left(
\begin{array}{ccc}
0 & 0 & (B_T/2M) \, {\rm sin} \, \phi_n \\
0& 0 & (B_T/2M) \, {\rm cos} \, \phi_n \\
(B_T/2M) \, {\rm sin}  \, \phi_n & (B_T/2M) \, {\rm cos} \, \phi_n & 0 \\
\end{array}
\right)~. 
\end{equation}
Denoting as in Section II-B by ${\cal U}_n (E,\phi_n)$ the associated transfer matrix with $y - y_0 = 150 \, {\rm pc}$, the transfer matrix describing the beam propagation over region 3 just follows from Eq. (\ref{t9nK26a}), namely
\begin{equation}
\label{as1g}
{\cal U}_3 \left(R_{\rm host}, R_*; E; \phi_1, . . . , \phi_{N_3} \right) = \prod^{N_3}_{n = 1} \, {\cal U}_n (E; \phi_n)~.
\end{equation}

We should however remark that such an effect plays a very minor role, and even ignoring it the final result is practically unaffected. This means that the back conversions $a \to \gamma$ effectively take place in region 2.

\subsection{Photon-ALP oscillations in extragalactic space}

What remains to be done is to address the beam propagation in extragalactic space, which is referred to as our region 4. 

We start by restricting our attention to conventional physics. As already pointed out, the diffuse infrared/optical/ultraviolet radiation generated by stars during the whole evolution of the Universe forms the extragalactic background light (EBL). Hence the beam photons can scatter off EBL photons thereby disappearing through the process $\gamma \gamma \to e^+ e^-$. As a consequence, the photon survival probability is usually expressed as
\begin{equation} 
\label{a012122010}
P_{\gamma \to \gamma}^{\rm CP} (E_0,z)  = e^{- \tau_{\gamma}(E_0,z)}~,
\end{equation}
where $\tau_{\gamma}(E_0,z)$ is the optical depth for the process $\gamma \gamma \to e^+ e^-$, which quantifies the dimming of a blazar at redshift $z$ observed at energy $E_0$. Hence the observed photon flux $F_{\rm obs} (E_0, z)$ is related to the intrinsic flux produced by the blazar $F_{\rm int} (E)$ through the relation
\begin{equation} 
\label{a02122010A}
F_{\rm obs}(E_0,z) = P_{\gamma \to \gamma}^{\rm CP} (E_0,z) \, F_{\rm em} \left( E \right) = e^{- \tau_{\gamma}(E_0,z)} \, F_{\rm em} \left( E \right)~,
\end{equation}
where of course $E = E_0 (1 + z)$. The optical depth is~\cite{pairproduction2}
\begin{eqnarray}
\label{eq:tau}
\tau_{\gamma}(E_0, z) = \int_0^{z} {\rm d} z^{\prime} ~ \frac{{\rm d} l(z^{\prime})}{{\rm d} z^{\prime}} \, \int_{-1}^1 {\rm d} {\mu} ~ \frac{1- \mu}{2} \ \times \\     \nonumber
\times  \, \int_{\epsilon_{\rm thr}(E(z^{\prime}) ,\mu)}^{\infty}  {\rm d} \epsilon (z^{\prime}) \, n_{\gamma} \bigl(\epsilon (z^{\prime}), z^{\prime} \bigr) \,  \sigma_{\gamma \gamma} \bigl( E(z^{\prime}), \epsilon(z^{\prime}), \mu \bigr)~, \ \ 
\end{eqnarray}
with $\epsilon (z) = \epsilon_0 (1 + z)$, where $\epsilon (z)$ denotes the energy of an EBL photon at redshift $z$ ($\epsilon_0$ is $\epsilon (z)$ at $z = 0$) while $n_{\gamma} \bigl(\epsilon (z), z \bigr)$ is the spectral number density of the EBL at redshift $z$. The quantities $\sigma_{\gamma \gamma} \bigl( E(z), \epsilon(z), \mu \bigr)$ and $\epsilon_{\rm thr} \bigl(E(z) ,\mu \bigr)$ have been defined by Eqs. (\ref{eq.sez.urto}) and (\ref{eq.sez.urto01012011}), respectively (we recall that $\mu$ is the cosine of the scattering angle). It can be shown that the considered cross-section gets maximized for
\begin{equation} 
\label{eq.sez.urto-1}
\epsilon_0 (E_0) \simeq \left(\frac{500 \, {\rm GeV}}{E_0 \left(1 + z \right)^2} \right) \, {\rm eV}~,
\end{equation}
where head-on collisions have been assumed for definiteness. This shows that the EBL is indeed of crucial importance for VHE astrophysics.

Finally, in the standard $\Lambda$CDM cosmological model the distance travelled by a photon per unit redshift at redshift $z$ is given 
by
\begin{equation}
\label{lungh}
\frac{d l(z)}{d z} = \frac{c}{H_0} \frac{1}{\left(1 + z \right) \left[ {\Omega}_{\Lambda} + {\Omega}_M \left(1 + z \right)^3 \right]^{1/2}}~,
\end{equation}
with $H_0 \simeq 70 \, {\rm Km} \, {\rm s}^{-1} \, {\rm Mpc}^{-1}$, ${\Omega}_{\Lambda} \simeq 0.7$ and ${\Omega}_M \simeq 0.3$. 

So, when $n_{\gamma} \bigl(\epsilon (z), z \bigr)$ is known, $\tau_{\gamma}(E_0, z)$ can be computed by performing a numerical integration of its above expression. Several realistic models for the EBL are available in the literature which rely on different strategies (see e.g.~\cite{ebl,eblmodels,frv,dominguez}). A remarkable fact is that they are basically in agreement with each other. Among all  possible choices, we will employ both the Franceschini-Rodighiero-Vaccari model~\cite{frv} and the Dominguez et al. model~\cite{dominguez}, which give practically identical results. 

\begin{figure}[t]
\centering
\vspace{-2 truecm}
\includegraphics[width=.55\textwidth]{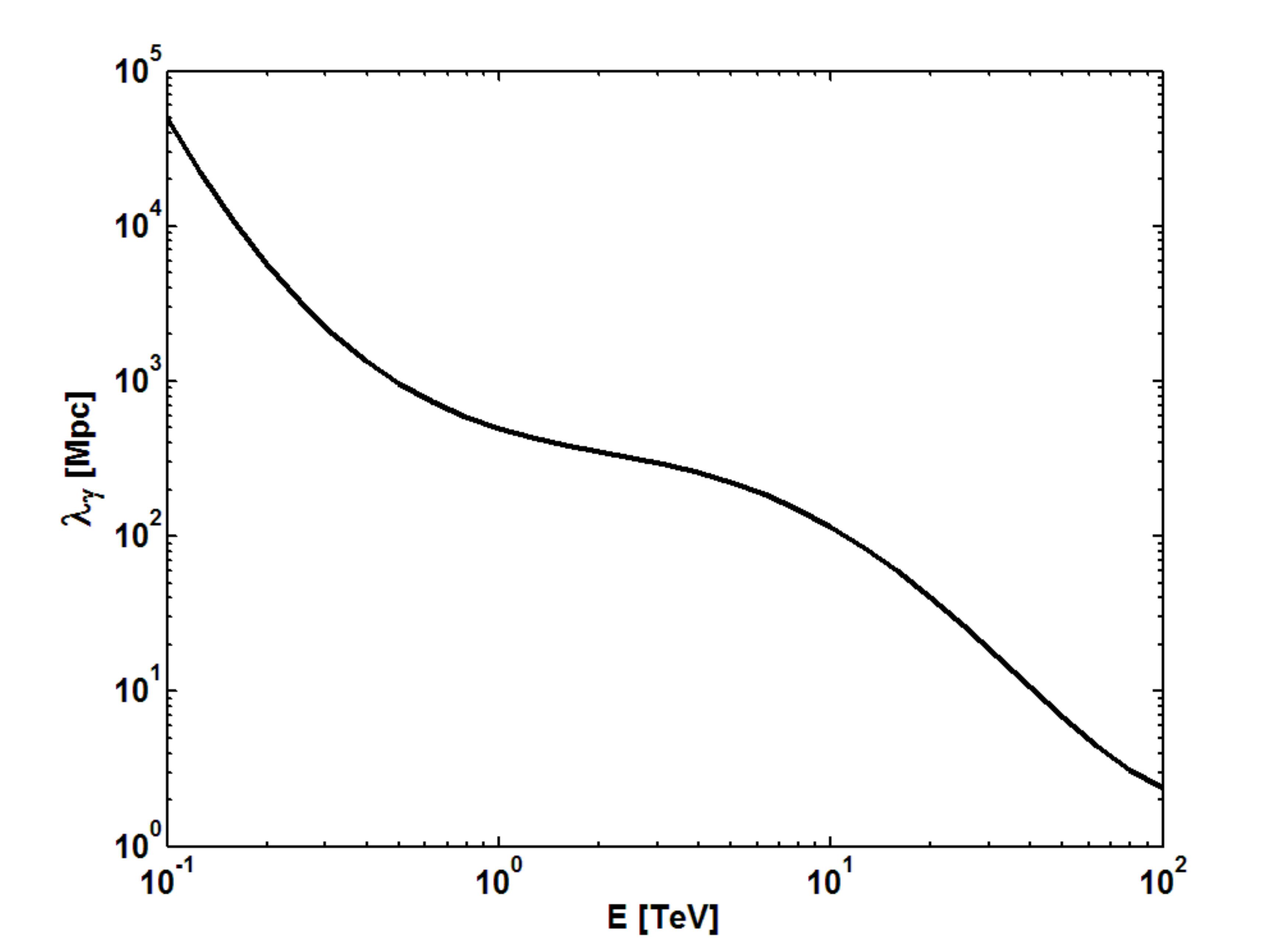}
\caption{\label{fig:france} 
The pair-production mean free path ${\lambda}_{\gamma}$ of a VHE photon is plotted versus its energy $E$ within the EBL model of Franceschini, Rodighiero and Vaccari~\cite{frv}. Only conventional physics is assumed and in particular the possibility of photon-ALP oscillations is ignored.}
\end{figure}

In order to get a feeling about the dimming of a blazar caused by the EBL absorption we discard cosmological effects so that the photon mean free path is
\begin{equation}
\label{odmfp}
\lambda_{\gamma} (E) = \frac{D}{\tau (E)}~,
\end{equation}
where $D$ is the distance to the blazar. We have plotted $\lambda_{\gamma} (E)$ versus $E$ in Fig. \ref{fig:france}.

Let us now consider the possibility that photon-ALP oscillations take place in extragalactic space according to the DARMA scenario,  which has been carefully discussed in~\cite{dgr}. Here we summarize the key points in order to make the present paper self-contained.

The crucial issue to address in this respect concerns the large-scale magnetic fields traversed by the beam, because we know that their presence is necessary for photon-ALP conversions to occur. Unfortunately, the origin and structure of these magnetic fields is still unknown to a large extent. 

A possibility is that very small magnetic fields present in the early Universe were subsequently amplified by the process of structure formation~\cite{dolag}. An alternative option is that the considered magnetic fields have been generated in the low-redshift Universe by energetic quasar outflows~\cite{furlanetto}. It has also been suggested that large-scale magnetic fields originated from the so-called Biermann battery effect~\cite{biermann}, namely from electric currents driven by merger shocks during the structure formation processes. Presumably, all these mechanisms can take place, even if it is presently impossible to assess their relative importance~\cite{lsmf}. At any rate, we suppose that magnetic fields already exist out to redshift $z \simeq 1$. We remark that the beam propagation is particularly sensitive to cosmic voids, and since the line of sight to PKS 1222+216 does not cross any cluster of galaxies, their magnetic field is irrelevant for the present purposes.

Regretfully, almost nothing is known about the morphology of large-scale magnetic fields. Even though it is fairly evident that their coherence length cannot be arbitrarily large, no reliable estimate of its value is available. This means that we cannot suppose that large-scale magnetic fields are homogeneous over the whole distance to the source, but their spatial dependence is largely unknown. The usual way out of this difficulty amounts to suppose that large-scale magnetic fields ${\bf B}$ have a {\it domain-like structure}. That is, ${\bf B}$ is assumed to be homogeneous over a domain of size $L_{\rm dom}$ roughly equal to its coherence length, with ${\bf B}$ randomly changing its direction from one domain to another but keeping approximately the same strength (more about this, later)~\cite{lsmf}. Therefore we can apply again the discussion in Section II-B with $\psi_n \to \varphi_n$ and $N \to N_4$, but in the cosmological setting it looks more natural to describe the overall domain-like network of the large-scale magnetic field by a uniform mesh in redshift space rather than in ordinary space. This can be done as follows. Denote by $n = 1$ the domain closest to us, which extends from $z = 0$ to $z = \Delta z$. Thanks to the linear Hubble law, we find
\begin{equation}
\label{mag2ZH}
\Delta z = 1.17 \cdot 10^{- 3} \left( \frac{L_{\rm dom}^{(1)} }{5 \, {\rm Mpc}} \right)~,
\end{equation}
where $L_{\rm dom}^{(1)}$ is the proper size of the first domain. But since all domains have the same redshift size by construction, it follows that their number $N_4$ is
\begin{equation}
\label{mag2ZI}
N_4 = \frac{z}{\Delta z} = 8.5 \cdot 10^2 \left( \frac{5 \, {\rm Mpc}}{L_{\rm dom}^{(1)} } \right) \, z~,
\end{equation}
which yields 
\begin{equation}
\label{mag2ZI}
N_4 = 3.67 \cdot 10^2 \left( \frac{5 \, {\rm Mpc}}{L_{\rm dom}^{(1)} } \right)
\end{equation}
for PKS 1222+216. Moreover, the proper size $L^{(n)}$ of the $n$-th domain can be evaluated by means of Eq. (\ref{lungh}) as
\begin{equation}
\label{lungh14012011}
L^{(n)}_{\rm dom} = \int_{(n - 1) \Delta z}^{n \Delta z} d z^{\prime} ~ \frac{d l(z^{\prime})}{d z^{\prime}} = 4.29 \cdot 10^3 \int_{(n - 1) \Delta z}^{n \Delta z} \frac{d z^{\prime}}{\left(1 + z^{\prime} \right) \left[ 0.7 + 0.3 \left(1 + z^{\prime} \right)^3 \right]^{1/2}} \ {\rm Mpc}~,
\end{equation}
which in the present situation approximately gives
\begin{equation}
\label{lunghK}
L^{(n)}_{\rm dom} = \frac{4.29 \cdot 10^3 \Delta z}{1 + 1.45 \, (n - 1) \Delta z} \, {\rm Mpc}~.
\end{equation}

The next piece of information that we need is the mean free path ${\lambda}^{(n)}_{\gamma}$ across the $n$-th domain. Applying  Eq. (69) to this domain and writing the integration over the interval $\bigl((n - 1) \Delta z, n \Delta z \bigr)$ as the integral from $0$ to $n \Delta z$ minus the integral from $0$ to $(n - 1) \Delta z$ we get $\tau_{\gamma} \left(E_0, n  \Delta z \right) - \tau_{\gamma} \bigl(E_0, (n - 1) \Delta z \bigr)$ for the optical depth of the domain in question. Now, because we have $\Delta z \sim 10^{- 3}$ cosmological effects are totally unimportant inside a single domain and so because of Eq. (\ref{odmfp}) we get
\begin{equation} 
\label{a02122010Aq}
{\lambda}^{(n)}_{\gamma} = \frac{L_{\rm dom}^{(n)}}{\tau_{\gamma} \left(E_0, n \, \Delta z \right) - \tau_{\gamma} \bigl(E_0, (n - 1) \Delta z \bigr)}~,
\end{equation}
which owing to Eq. (\ref{lunghK}) takes the form 
\begin{equation} 
\label{a02122010Ar}
{\lambda}^{(n)}_{\gamma}  = \left( \frac{4.29 \cdot 10^3}{1 + 1.45 \, (n - 1) \Delta z} \right) \left( \frac{\Delta z }{\tau_{\gamma} \left(E_0, n \, \Delta z \right) - \tau_{\gamma} \bigl(E_0, (n - 1) \Delta z \bigr)} \right) \, {\rm Mpc}~.
\end{equation}

At variance with the case treated in Section III-D, $B$ is not exactly the same in all domains. For, the Gunn Peterson effect~\cite{gunnpeterson} tells us that the Universe is ionized and owing to the high conductivity of the extragalactic medium the magnetic flux lines can be thought as frozen inside it. Hence, flux conservation during the cosmic expansion implies that $B$ scales like the volume to the power $2/3$, thereby entailing that the magnetic field strength $B^{(n)}_T$ in the $n$-th magnetic domain is~\cite{lsmf}  
\begin{equation}
\label{mag1}
B^{(n)}_T = B_{T,0} \, \bigl[(1 + (n -1) \Delta z \bigr]^2~,       
\end{equation}
with $B_{T,0}$ denoting its value at $z = 0$.

Finally, the mean diffuse extragalactic electron density obeys the constraint $n_e < 2.7 \cdot 10^{- 7} \, {\rm cm}^{- 3}$ arising from  the WMAP measurement of the baryon density~\cite{wmap}, but it has been argued that for $z < 1$ such a bound can be stronger by a factor 15~\cite{csaki2}. 

As shown elsewhere~\cite{dgr}, the effect of photon-ALP oscillations in extragalactic space is maximized for a large-scale magnetic field with $B$ in the nG range and $L_{\rm dom}$ in the Mpc range -- which is consistent with all observational constraints~\cite{lsmf} and with the AUGER results~\cite{auger} -- provided that the strong-mixing regime is realized. The DARMA scenario assumes that this is indeed the case for $E_L \simeq 100 \, {\rm GeV}$, but we can just as well require $E_L \simeq 70 \, {\rm GeV}$ (the QED one-loop effect is here ridiculously small). Correspondingly, recalling Eq. (\ref{t9nK}) the following constraint has to be met
\begin{equation}
\label{t9nKQ}
\left| \left(\frac{m}{10^{- 10} \, {\rm eV}} \right)^2 - 0.13 \left(\frac{n_e}{{\rm cm}^{- 3}} \right) \right| \left(\frac{nG}{B_T} \right) \left(\frac{M}{10^{11} \, {\rm GeV}} \right) < 2.8~. 
\end{equation}
Under such an assumption, the plasma contribution and the ALP mass term should be dropped and from Eqs. (\ref{t2qq1}), (\ref{t2qq2}), (\ref{t2qq4}) and (\ref{t2qq5}) we find for the $\Delta$ terms the following expressions
\begin{equation}
\label{t2qq1X} 
\Delta^{(n)}_{\bot} (E_0) = \frac{i}{2 \, {\lambda}^{(n)}_{\gamma}(E_0)}~,
\end{equation}
\begin{equation}
\label{t2qq2X} 
\Delta^{(n)}_{\parallel} (E_0) = \frac{i}{2 \, {\lambda}^{(n)}_{\gamma}(E_0)}~,    
\end{equation}
\begin{equation}
\label{t2qq4X} 
\Delta^{(n)}_{a \gamma}  = \frac{B^{(n)}_T}{2 M}~,
\end{equation}
\begin{equation}
\label{t2qq5X} 
\Delta^{(n)}_{a a} (E_0) = 0~,
\end{equation}
thereby implying that in the $n$-th generic domain ($1 \leq n \leq N_4$) the mixing matrix (\ref{t3G}) takes the explicit form
\begin{equation}   
\label{t3GN}
{\cal M}_n (E_0, \varphi_n) = \left(
\begin{array}{ccc}
\frac{i}{2 \, {\lambda}^{(n)}_{\gamma}(E_0)} & 0 & \frac{B^{(n)}_T}{2 M} \, {\rm sin} \, \varphi_n \\
0& \frac{i}{2 \, {\lambda}^{(n)}_{\gamma}(E_0)} & \frac{B^{(n)}_T}{2 M} \, {\rm cos} \, \varphi_n \\
\frac{B^{(n)}_T}{2 M} \, {\rm sin}  \, \varphi_n & \frac{B^{(n)}_T}{2 M} \, {\rm cos} \, \varphi_n & 0 \\
\end{array}
\right)~. 
\end{equation}
After some tedious algebra, following the discussion in Section II the transfer matrix associated with the $n$-th domain is found to be
\begin{equation}
\label{mravvq2abcQQ}
{\cal U}_n (E_0, \varphi_n) = e^{i \left({\lambda}^{(n)}_1 \, L_{\rm dom}^{(n)} \right)} \, T_1 (\varphi_n) + e^{i \left({\lambda}^{(n)}_2  \, L_{\rm dom}^{(n)} \right)} \, T_2 (\varphi_n) + e^{i \left({\lambda}^{(n)}_3  \, L_{\rm dom}^{(n)} \right)} \, T_3 (\varphi_n)~, 
\end{equation}
with
\begin{equation}
\label{mravvq2Q1a}
T_1 (\varphi_n) \equiv
\left(
\begin{array}{ccc}
\cos^2 \varphi_n & -\sin \varphi_n \cos \varphi_n & 0 \\
- \sin \varphi_n \cos \varphi_n & \sin^2 \varphi_n & 0 \\
0 & 0 & 0
\end{array}
\right)~,
\end{equation}
\begin{equation}
\label{mravvq3Q1b}
T_2 (\varphi_n) \equiv 
\left(
\begin{array}{ccc}
\frac{- 1 + \sqrt{1 - 4 {\delta}_n^2}}{2 \sqrt{1 - 4 {\delta}_n^2}} \sin^2 \varphi_n & \frac{- 1 + \sqrt{1 - 4 {\delta}_n^2}}{2 \sqrt{1 - 4 {\delta}_n^2}} \sin \varphi_n \cos \varphi_n & \frac{i \delta_n}{\sqrt{1 - 4 {\delta}_n^2}} \sin \varphi_n \\
\frac{- 1 + \sqrt{1 - 4 {\delta}_n^2}}{2 \sqrt{1 - 4 {\delta}_n^2}} \sin \varphi_n \cos \varphi_n & \frac{- 1 + \sqrt{1 - 4 {\delta}_n^2}}{2 \sqrt{1 - 4 {\delta}_n^2}} \cos^2 \varphi_n & \frac{i \delta_n}{\sqrt{1 - 4 {\delta}_n^2}} \cos \varphi_n \\
\frac{i \delta_n}{\sqrt{1 - 4 {\delta}_n^2}} \sin \varphi_n & \frac{i \delta_n}{\sqrt{1 - 4 {\delta_n}^2}} \cos \varphi_n & \frac{ 1 + \sqrt{1 - 4 {\delta}_n^2}}{2 \sqrt{1 - 4 {\delta}_n^2}}
\end{array}
\right)~,
\end{equation}
\begin{equation} 
\label{mravvq2Q1b}
T_3 (\varphi_n) \equiv
\left(
\begin{array}{ccc}
\frac{ 1 + \sqrt{1 - 4 {\delta}_n^2}}{2 \sqrt{1 - 4 {\delta}_n^2}} \sin^2 \varphi_n  & \frac{ 1 + \sqrt{1 - 4 {\delta}_n^2}}{2 \sqrt{1 - 4 {\delta}_n^2}} \sin \varphi_n \cos \varphi_n  & \frac{- i \delta_n}{\sqrt{1 - 4 {\delta}_n^2}}    \sin \varphi_n \\ 
\frac{ 1 + \sqrt{1 - 4 {\delta}_n^2}}{2 \sqrt{1 - 4 {\delta}_n^2}} \sin \varphi_n \cos \varphi_n  & \frac{ 1 + \sqrt{1 - 4 {\delta}_n^2}}{2 \sqrt{1 - 4 {\delta}_n^2}} \cos^2 \varphi_n  & \frac{- i \delta_n}{\sqrt{1 - 4 {\delta}_n^2}} \cos \varphi_n \\
\frac{- i \delta_n}{\sqrt{1 - 4 {\delta}_n^2}} \sin \varphi_n  & \frac{- i \delta_n}{\sqrt{1 - 4 {\delta}_n^2}} \cos \varphi_n  &  \frac{- 1 + \sqrt{1 - 4 {\delta}_n^2}}{2 \sqrt{1 - 4 {\delta}_n^2}}   
\end{array}
\right)~,
\end{equation} 
where the eigenvalues of ${\cal M}_n (E_0, \varphi_n)$ are~\cite{comment31}
\begin{equation}
\label{a91212a1PW}
{\lambda}^{(n)}_{1} (E_0) = \frac{i}{2 \, {\lambda}^{(n)}_{\gamma} (E_0)}~,
\end{equation}
\begin{equation}
\label{a91212a2PW}
{\lambda}^{(n)}_{2} (E_0) = \frac{i}{4 \, {\lambda}^{(n)}_{\gamma}(E_0)} \left(1 - \sqrt{1 - 4 \, \delta^2_n} \right)~, 
\end{equation}
\begin{equation}
\label{a91212a3PW}
{\lambda}^{(n)}_{3} (E_0) = \frac{i}{4 \, {\lambda}^{(n)}_{\gamma}(E_0)}  \left(1 + \sqrt{1 - 4 \, \delta^2_n} \right)~,
\end{equation}
and we have set for notational simplicity
\begin{equation}
\label{a17PW}
{\delta}_n \equiv \frac{B^{(n)}_T \, {\lambda}^{(n)}_{\gamma} (E_0)}{M}~.
\end{equation}
Thus, the transfer matrix describing the beam propagation over region 4 just follows from Eq. (\ref{t9nK26a}), namely
\begin{equation}
\label{as1g}
{\cal U}_4 \left(D,R_{\rm host}; E_0; \varphi_1, . . . , \varphi_{N_4} \right) = \prod^{N_4}_{n = 1} \, {\cal U}_n (E_0, \varphi_n)~,
\end{equation}
where $D$ denotes the distance of PKS 1222+216.

\section{RESULTS}

Let us now focus our attention on the overall effect of ALPs on the beam propagation from the central region of PKS 1222+216 to the Earth. This amounts to compute the photon survival probability $P^{\rm ALP}_{\gamma \to \gamma}(E)$, which allows us in turn to figure out how the intrinsic VHE blazar emitted spectrum looks like. We will consider separately the cases in which photon-ALP occur or not in extragalactic space.

\subsection{No oscillations in extragalactic space}

We first suppose that for whatever reason (too small large-scale magnetic fields, too large ALP mass, an so on) the effect of photon-ALP oscillations in extragalactic space is negligible or even absent altogether. In this way, we can concentrate ourselves on the 
relevance of $\gamma \to a$ and $a \to \gamma$ conversions inside the source and around it. 

\begin{figure}[h]
\centering
\vspace{-2 truecm}
\includegraphics[width=.55\textwidth]{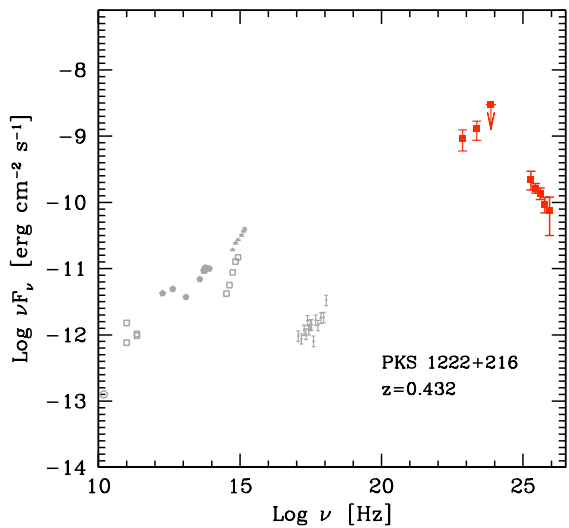}
\vspace{-2 truecm}
\caption{\label{fig:sedobs} Red points at high and VHE are the spectrum of PKS 1222+216 recorded by {\it Fermi}/LAT and the one detected by MAGIC but EBL-deabsorbed according to conventional physics using Eq. (\ref{a02122010Am18d}).}
\end{figure}

Now, in the first place we EBL-deabsorb the observed flux $F_{\rm obs}(E_0,z)$ obtained by MAGIC as usual, so as to get the flux at the edge of the host galaxy. Owing to Eq. (\ref{a02122010A}), we find
\begin{equation} 
\label{a02122010Am18d}
F_{\rm R_{\rm host}} \left(E \right) = \frac{F_{\rm obs}(E_0,z)}{P_{\gamma \to \gamma}^{\rm CP} (E_0,z)} = e^{\tau_{\gamma}(E_0,z)}  \, F_{\rm obs}(E_0,z)~,
\end{equation}
with $E = E_0 (1 + z)$, where $E_0$ denotes the observed energy and $z = 0.432$ for PKS 1222+216. Manifestly, $F_{\rm R_{\rm host}} \left(E \right)$ would be the intrinsic emitted flux $F_{\rm em} \left(E \right)$ in the absence of BLR absorption. We exhibit $F_{\rm R_{\rm host}} \left(E \right)$ in Fig. \ref{fig:sedobs} together with the observed {\it Fermi}/LAT spectrum.

Since we know the transfer matrices ${\cal U}_1 (R_{\rm BLR}, 0;E)$, ${\cal U}_2 (R_*, R_{\rm BLR};E)$ and 
${\cal U}_3 \left(R_{\rm host}, R_*; E; \phi_1, . . . , \phi_{N_3} \right)$, we can apply the results of our general discussion presented in Section II.

Recalling the definitions of ${\cal U}_{\rm smooth} (E)$ and ${\cal U}_{\rm random} \left(E; \psi_1, ... , \psi_N \right)$ just after Eq. (\ref{t9nK26a}), we presently have
\begin{equation}
\label{t2qq5M18a} 
{\cal U}_{\rm smooth} (E) = {\cal U}_2 (R_*, R_{\rm BLR};E) \,  {\cal U}_1 (R_{\rm BLR}, 0;E)~,
\end{equation}
\begin{equation}
\label{t2qq5M18b} 
{\cal U}_{\rm random} \left(E; \phi_1, ... , \phi_{N_3} \right) = {\cal U}_3 \left(R_{\rm host}, R_*; E; \phi_1, . . . , \phi_{N_3} \right)~,
\end{equation}
in terms of which we can use Eq. (34) to get the explicit expression of Eq. (\ref{t7280527b}) and ultimately of Eq. (39), namely
\begin{eqnarray}
\label{k3lwf1w1WWm18c}
P^{\rm ALP}_{\gamma \to \gamma} (R_{\rm host},0;E)  =  \Big\langle P_{\rho_{\rm unpol} \to \rho_x} \left(R_{\rm host},0;E; \phi_1, ... , \phi_{N_3} \right) \Big\rangle_{\phi_1, ... , \phi_{N_3}} +   \\  \nonumber
+ \, \Big\langle P_{\rho_{\rm unpol} \to \rho_z} \left(R_{\rm host},0;E; \phi_1, ... , \phi_{N_3} \right) 
\Big\rangle_{\phi_1, ... , \phi_{N_3}}~.   \ \ \ \ \ \ \ \ \ \ \ 
\end{eqnarray}
Therefore, in terms of $F_{\rm R_{\rm{host}}} \left(E \right)$ as dictated by Eq. (\ref{a02122010Am18d}), according to our model $F_{\rm em} \left(E \right)$ is given by
\begin{equation}
F_{\rm em} \left(E \right) = \frac{F_{\rm R_{\rm{host}}} \left(E \right)}{P^{\rm ALP}_{\gamma \to \gamma} \left(R_{\rm host}, 0; E \right)}~.
\label{abc0606}
\end{equation}

Manifestly, what we have to do at this stage is to make a choice for $M$ (consistent with the bound $M > 1.14 \cdot 10^{10}$ GeV set by CAST), and even though we have fixed the magnetic field inside the source before the BLR at $B=0.2$ G this value is uncertain to some extent. 

Since the principal motivation of the present work is to explain why a substantial fraction of VHE photons escape from the BLR, we feel that a deeper insight into the suitable values of $M$ and $B$ can be gained by addressing the effective optical depth $\tau_{\rm eff} (E)$ concerning the photon propagation from the central region to the edge of the host galaxy, which is defined in terms of $P^{\rm ALP}_{\gamma \to \gamma} \left(R_{\rm host}, 0 ; E \right)$ in complete analogy with Eq. (\ref{tM}) as
\begin{equation} 
\label{a02122010Am18e}
P^{\rm ALP}_{\gamma \to \gamma} \left(R_{\rm host}, 0 ; E \right) = e^{- \tau_{\rm eff} (E)}~.
\end{equation}
After some attempts, we have been led to select for definiteness three benchmark cases: $(B = 0.2 \, {\rm G}, M = 7 \cdot 10^{10} \, {\rm GeV})$, $(B = 0.4 \, {\rm G}, M = 1.5 \cdot 10^{11} \, {\rm GeV})$ and $(B = 2 \, {\rm G}, M = 4 \cdot 10^{11} \, {\rm GeV})$. We report the corresponding curves for $\tau_{\rm eff} (E)$ in Fig. \ref{fig:surf}, where the red solid line corresponds to $(B = 0.2 \, {\rm G}, M = 7 \cdot 10^{10} \, {\rm GeV})$, the green short-dashed line to $(B = 0.4 \, {\rm G}, M = 1.5 \cdot 10^{11} \, {\rm GeV})$ and the violet dashed-dotted line to $(B = 2 \, {\rm G}, M = 4 \cdot 10^{11} \, {\rm GeV})$, while the blue long-dashed line corresponds to conventional physics (as in~\cite{tavecchiomazin2009}). The effect of the photon-ALP oscillations on the beam propagation can readily be appreciated. Indeed, photon-ALP oscillations lead to a drastic reduction of the optical depth in the optically thick range. Our best case in this respect is $(B = 0.2 \, {\rm G}, M = 7 \cdot 10^{10} \, {\rm GeV})$, where in the MAGIC band the effective optical depth is almost constant at about $\tau _{\rm eff} \simeq 4$, corresponding to a survival probability of about $2\%$. On the contrary, in the optically thin region below $\sim \, 30 \, {\rm GeV}$ the optical depth in the presence of photon-ALP oscillations is {\it larger} than the standard one, which instead goes rapidly to zero below 10 GeV. This behaviour can be simply understood: A fraction around $10 \%$ of the $\gamma$ rays originally emitted by the source and converted into ALPs do not reconvert back to photons, therefore leading to a reduction of the observed  photon flux.

\begin{figure}[h]
\centering
\includegraphics[width=.55\textwidth]{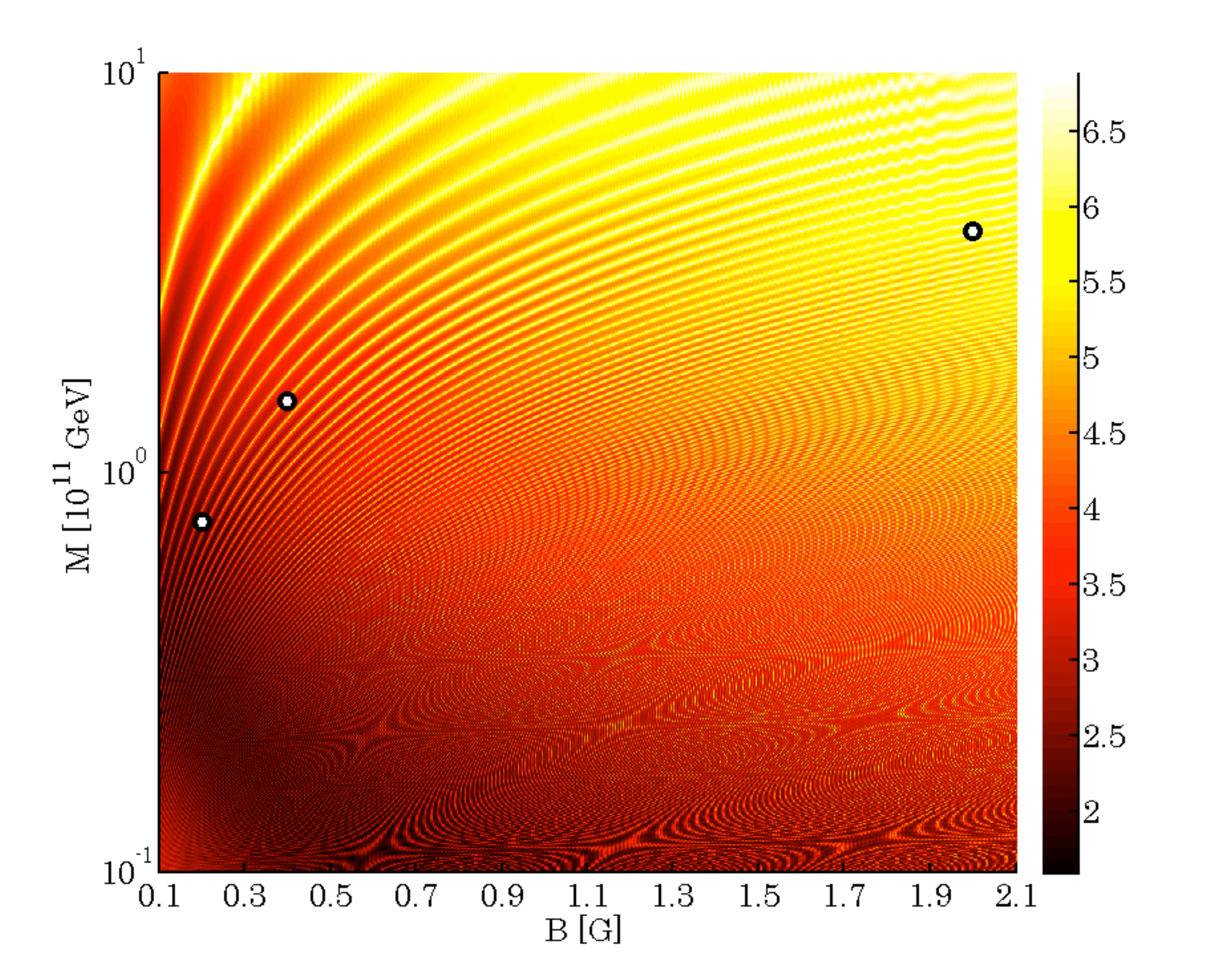}
\caption{\label{fig:probgrande} The quantity $\Pi$ as a function of $B$ and $M$. The three white blobs correspond to our benchmark cases.}
\end{figure}

Still, our goal is not merely to explain why MAGIC has observed PKS 1222+216 but also to find a realistic and physically motivated  SED that fits both {\it Fermi}/LAT and MAGIC spectra. So, it is not enough that photon-ALP oscillations allow for a large photon fraction to avoid the BLR absorption but they also have to give rise to a SED with the above features. In order to settle this issue we find it illuminating to proceed as follows. As we said, our source has been observed by {\it Fermi}/LAT in the energy range $0.3 - 3\, {\rm GeV}$ and by MAGIC in the band $70 - 400 \, {\rm GeV}$.  Therefore we focus our attention on the energies $E = 1 \, {\rm GeV}$ and $E = 300 \, {\rm GeV}$ as representative of the two kinds of measurements. Hence it is useful to define the ratio 
\begin{equation} 
\label{a02122010Am18f}
{\Pi} \equiv {\rm log} \left(\frac{P^{\rm ALP}_{\gamma \to \gamma} \left(R_{\rm host}, 0 ; 1 \, {\rm GeV} \right)}{P^{\rm ALP}_{\gamma \to \gamma} \left(R_{\rm host}, 0 ; 300 \, {\rm GeV} \right)} \right)~.
\end{equation}
A glance to Fig. \ref{fig:sedobs} shows that in order to have an acceptable shape for the emitted VHE component of the SED we need to have $\Pi$ as low as possible. Moreover, since the photon-ALP conversion is in any case small in the {\it Fermi}/LAT energy range, a small $\Pi$ would imply a huge correction of the flux in the MAGIC spectrum. Therefore, $\Pi$ allows us to have an effective control of the effects of the photon-ALP oscillations on the corrected SED. We show in Fig. \ref{fig:probgrande}  ${\Pi}$ as a function of $B$ and $M$ in order to find out how strongly ${\Pi}$ depends on these two parameters. Incidentally, the oscillatory behavior displayed by ${\Pi}$ in Fig. \ref{fig:probgrande} arises from $P_{\gamma \to \gamma} (1 \, {\rm GeV})$ which is in the strong-mixing and absorption-free regime, even though outside $R_{\rm BLR}$ ${\bf B}$ first decreases and then has a random domain-like structure. The oscillatory behavior makes the derived probability a rather complex function of the two parameters. As a general trend, low values of $\Pi$ are associated with low $B$ and $M$ values (lower left corner). On the contrary, large  $B$ and $M$ (upper right corner) result in large $\Pi$. Our three benchmark cases are represented by the three white blobs in Fig. \ref{fig:probgrande}.

Quite remarkably, we see that the trend found in Fig. \ref{fig:surf} is reproduced in Fig. \ref{fig:probgrande}, in the sense to a progressively increasing $\tau (E)$ there correspond higher values of $\Pi$. Thus, we are led to the conclusion that the case $(B = 0.2 \, {\rm G}, M = 7 \cdot 10^{10} \, {\rm GeV})$ is expected to have also a most realistic SED. Besides, also the cases $(B = 0.4 \, 
{\rm G}, M = 1.5 \cdot 10^{11} \, {\rm GeV}$ and $B = 2 \, {\rm G}, M = 4 \cdot 10^{11} \, {\rm GeV}$ look promising as far as the shape of the SED is concerned even if the corresponding behavior of $\tau _{\rm eff}(E)$ looks rather high especially at the highest energies, thereby implying a rather hard intrinsic spectrum. We shall come back to a thorough discussion of the SED in Section V.

\begin{figure}
\centering
\vspace{-2 truecm}
\includegraphics[width=.50\textwidth]{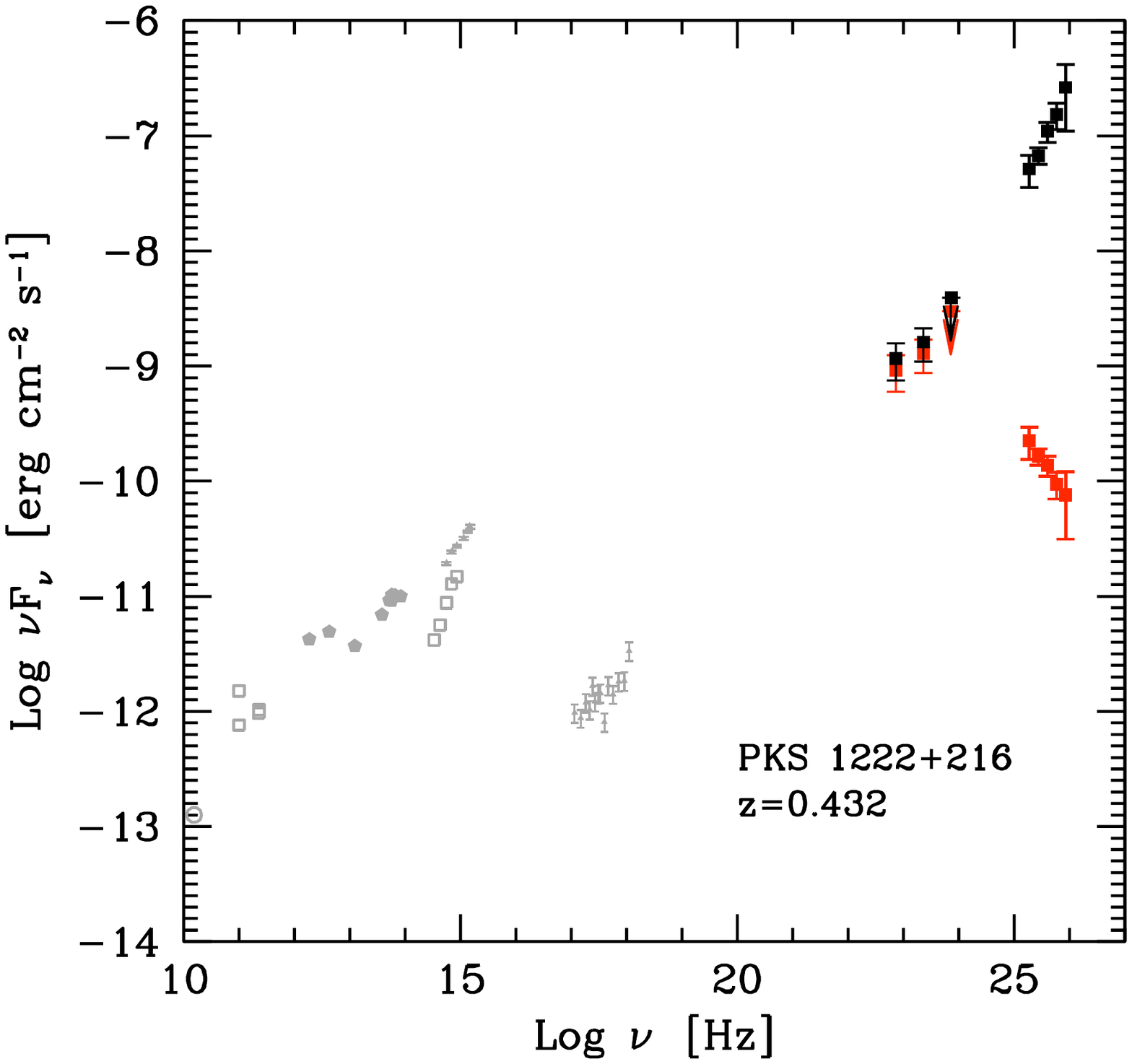}
\vspace{-2 truecm}
\caption{\label{fig:sedb} Red points at high energy and VHE are the spectrum of PKS 1222+216 recorded by {\it Fermi}/LAT and the one detected by MAGIC but EBL-deabsorbed according to conventional physics using Eq. (\ref{a02122010Am18d}). The black points represent the same data once further corrected for the photon-ALP oscillation effect employing Eq. (\ref{abc0606}) in the case $(B = 0.4 \, {\rm G}, M = 1.5 \cdot 10^{11} \, {\rm GeV})$. The gray data points below $10^{20} \, {\rm Hz}$ are irrelevant for the present discussion (details can be found in~\cite{tavecchioetal2011}.}
\end{figure}

Let us next explicitly address the impact of our model for the emitted spectrum of PKS 1222+216, which is shown in Figs. \ref{fig:seda} and \ref{fig:sedb} for the cases  $(B = 0.2 \, {\rm G}, M = 7 \cdot 10^{10} \, {\rm GeV})$ and $(B = 0.4 \, {\rm G}, M = 1.5 \cdot 10^{11} \, {\rm GeV})$, respectively, where the EBL-deabsorbed MAGIC points according to Eq. (\ref{a02122010Am18d}) and the observed {\it Fermi}/LAT points are reported in red whereas the black points are correspondingly obtained by means of Eq. (\ref{abc0606}). We do not report the Figure pertaining to the case $(B = 2 \, {\rm G}, M = 4 \cdot 10^{11} \, {\rm GeV})$ because the 
$\gamma$-ray peak would give an unacceptably large value for $\nu F_{\nu}$.

\subsection{Oscillations in extragalactic space}

We now consider the possibility that photon-ALP oscillations efficiently occur also in extragalactic space, which is possible only within the strong-mixing regime so that condition (\ref{t9nKQ}) has to be met. Let us consider first the case ($B = 0.2 \, {\rm G}, M = 7 \cdot 10^{10} \, {\rm GeV}$), assuming a domain-like large-scale magnetic field with strength $B = 0.7 \, {\rm nG}$ and coherence length $L_{\rm dom} = 4 \, {\rm Mpc}$, a situation corresponding to the most favorable case considered within the DARMA scenario~\cite{dgr}. 
\begin{figure}[b]
\centering
\vspace{-2 truecm}
\includegraphics[width=.50\textwidth]{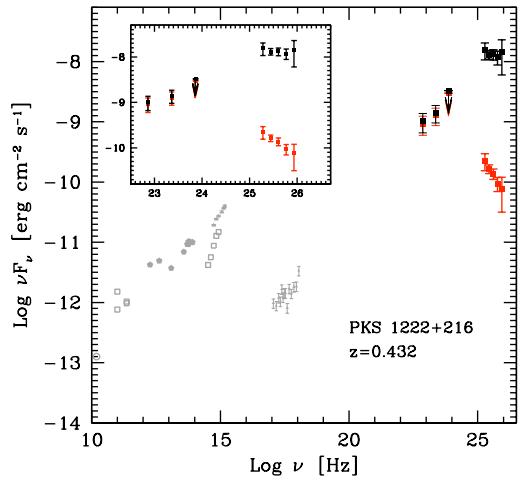}
\vspace{-2 truecm}
\caption{\label{fig:seda} Red points at high energy and VHE are the spectrum of PKS 1222+216 recorded by {\it Fermi}/LAT and the one detected by MAGIC but EBL-deabsorbed according to conventional physics using Eq. (\ref{a02122010Am18d}). The black points represent the same data once further corrected for the photon-ALP oscillation effect employing Eq. (\ref{abc0606}) in the case $(B = 0.2 \, {\rm G}, M = 7 \cdot 10^{10} \, {\rm GeV})$. The gray data points below $10^{20} \, {\rm Hz}$ are irrelevant for the present discussion (details can be found in~\cite{tavecchioetal2011}.}
\end{figure}~
~

\bigskip

\begin{figure}[h]
\centering
\vspace{-2 truecm}
\includegraphics[width=.50\textwidth]{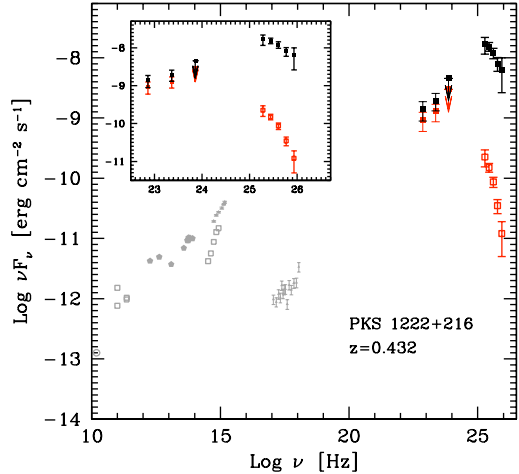}
\vspace{-2 truecm}
\caption{\label{fig:sedebl} Red points at high energy and open red squares at VHE are the spectrum of PKS 1222+216 recorded by {\it Fermi}/LAT and the one {\it observed} by MAGIC and not EBL-deabsorbed. The black points represent the same data once further corrected for the photon-ALP oscillation effect in the case $(B = 0.2 \, {\rm G}, M = 7 \cdot 10^{10} \, {\rm GeV})$ including also photon-ALP oscillations in extragalactic space, where a magnetic field with strength $B=0.7$ nG is supposed to exist. So, they are obtained from the red points and the open red squares by means of Eq. (\ref{abc0606Z}). The gray data points below $10^{20} \, {\rm Hz}$ are irrelevant for the present discussion (details can be found in~\cite{tavecchioetal2011}).}
\end{figure}

Accordingly, condition (\ref{t9nKQ}) is satisfied for $m < 1.7 \cdot 10^{- 10} \, {\rm eV}$, which is consistent with the upper bound found above. 

The form of ${\cal U}_{\rm smooth} (E)$ as given by Eq. (\ref{t2qq5M18a}) remains unchanged, but that of  ${\cal U}_{\rm random} \left(E; \psi_1, ... , \psi_N \right)$ now becomes
\begin{equation}
\label{t2qq5M18bz} 
{\cal U}_{\rm random} \left(E; \phi_1, ... , \phi_{N_3}; \varphi_1, ... , \varphi_{N_4} \right) = {\cal U}_4 \left(D,R_{\rm host}; E_0; \varphi_1, . . . , \varphi_{N_4} \right) \, {\cal U}_3 \left(R_{\rm host}, R_*; E; \phi_1, . . . , \phi_{N_3} \right)~,
\end{equation}
with $E = E_0 (1 + z)$. Just as before, we use these equations to find the explicit expression of Eq. (\ref{t7280527b}) and finally of Eq. (39), which presently reads
\begin{eqnarray}
\label{k3lwf1w1WWm18cZ}
P^{\rm ALP}_{\gamma \to \gamma} (D,0;E)  =  \Big\langle P_{\rho_{\rm unpol} \to \rho_x} \left(D,0;E; \phi_1, ... , \phi_{N_3}; \varphi_1, ... , \varphi_{N_4} \right) \Big\rangle_{\phi_1, ... , \phi_{N_3}; \varphi_1, ... , \varphi_{N_4}} +   \\  \nonumber
+ \, \Big\langle P_{\rho_{\rm unpol} \to \rho_z} \left(D,0;E; \phi_1, ... , \phi_{N_3}; \varphi_1, ... , \varphi_{N_4} \right) 
\Big\rangle_{\phi_1, ... , \phi_{N_3}; \varphi_1, ... , \varphi_{N_4}}~.   \ \ \ \ \ \ \ \ \ \ \ 
\end{eqnarray}
However, at variance with the previous treatment, the intrinsic flux emitted by the source -- which is represented in Fig. \ref{fig:sedebl} by black dots -- is obtained directly from the one {\it observed} by MAGIC and represented in Fig. \ref{fig:sedebl} by open red squares through the relation
\begin{equation}
F_{\rm em} (E) = \frac{F_{\rm obs} (E_0,z)}{P^{\rm ALP}_{\gamma \to \gamma} \left(D, 0; E \right)}~.  
\label{abc0606Z}
\end{equation}

The comparison of Figs. \ref{fig:seda} and \ref{fig:sedebl} reveals that the inclusion of the photon-ALP oscillations in extragalactic space does not lead to a dramatic effect. Although the derived intrinsic spectrum is softer, the peak energy and the luminosity of the high energy peak are roughly the same in either case. 

A very similar situation concerns the cases $(B = 0.4 \, {\rm G}, M = 1.5 \cdot 10^{11} \, {\rm GeV})$ and $(B = 2 \, {\rm G}, M = 4 \cdot 10^{11} \, {\rm GeV})$, and so we do not find it useful to report also these results in a Figure (in particular photon-ALP oscillations in extragalactic space do not save the case $(B = 2 \, {\rm G}, M = 4 \cdot 10^{11} \, {\rm GeV})$ so that it remains ruled out).

We conclude that in the context of our model for PKS 1222+216 photon-ALP oscillations in the extragalactic space are allowed but not compelling. 

\section{SPECTRAL ENERGY DISTRIBUTION (SED)}

Our final step consists in showing that the emitted spectra in the whole $\gamma$-ray band obtained in Section IV indeed 
lie on the SED of a realistic and physically motivated blazar model, thereby closing the circle. We stress that this problem does not have a unique solution, in the sense that it is quite conceivable that various leptonic and even hadronic models can work. Nevertheless, from a methodological point of view the present work would be incomplete without the presentation of an explicit emission model.

\begin{figure}[h]
\centering
\vspace{-2 truecm}
\includegraphics[width=.52\textwidth]{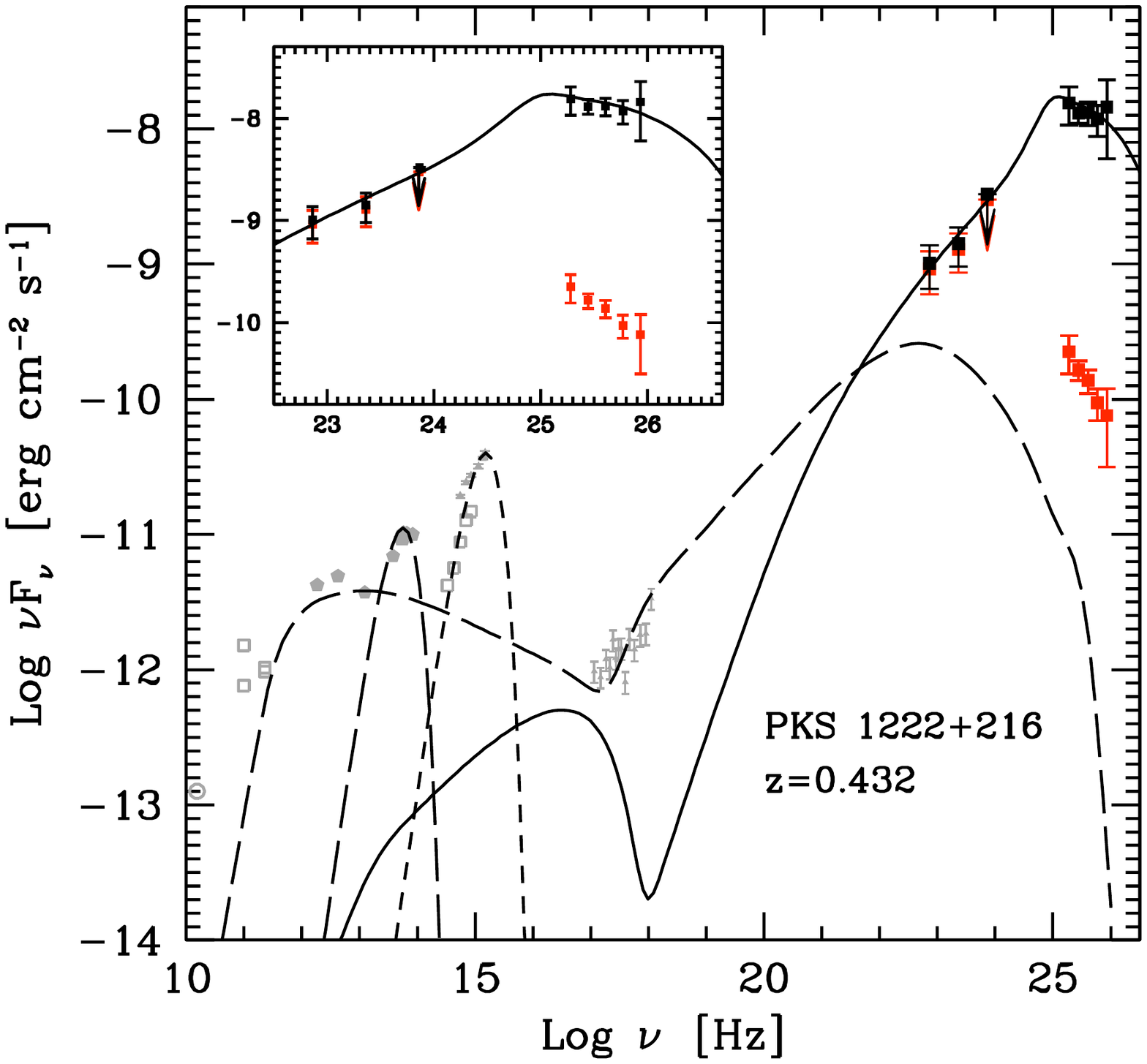}
\vspace{-2 truecm}
\caption{\label{fig:seda1} Same as Fig.~\ref{fig:seda} for the case $(B = 0.2 \, {\rm G}, M = 7 \cdot 10^{10} \, {\rm GeV})$ without photon-ALP oscillations in extragalactic space, but in addition the dashed and solid curves show the SED resulting from the considered two blobs which account for the $\gamma$-ray emission at high energy and VHE, respectively. }
\end{figure}

\begin{figure}
\centering
\includegraphics[width=.52\textwidth]{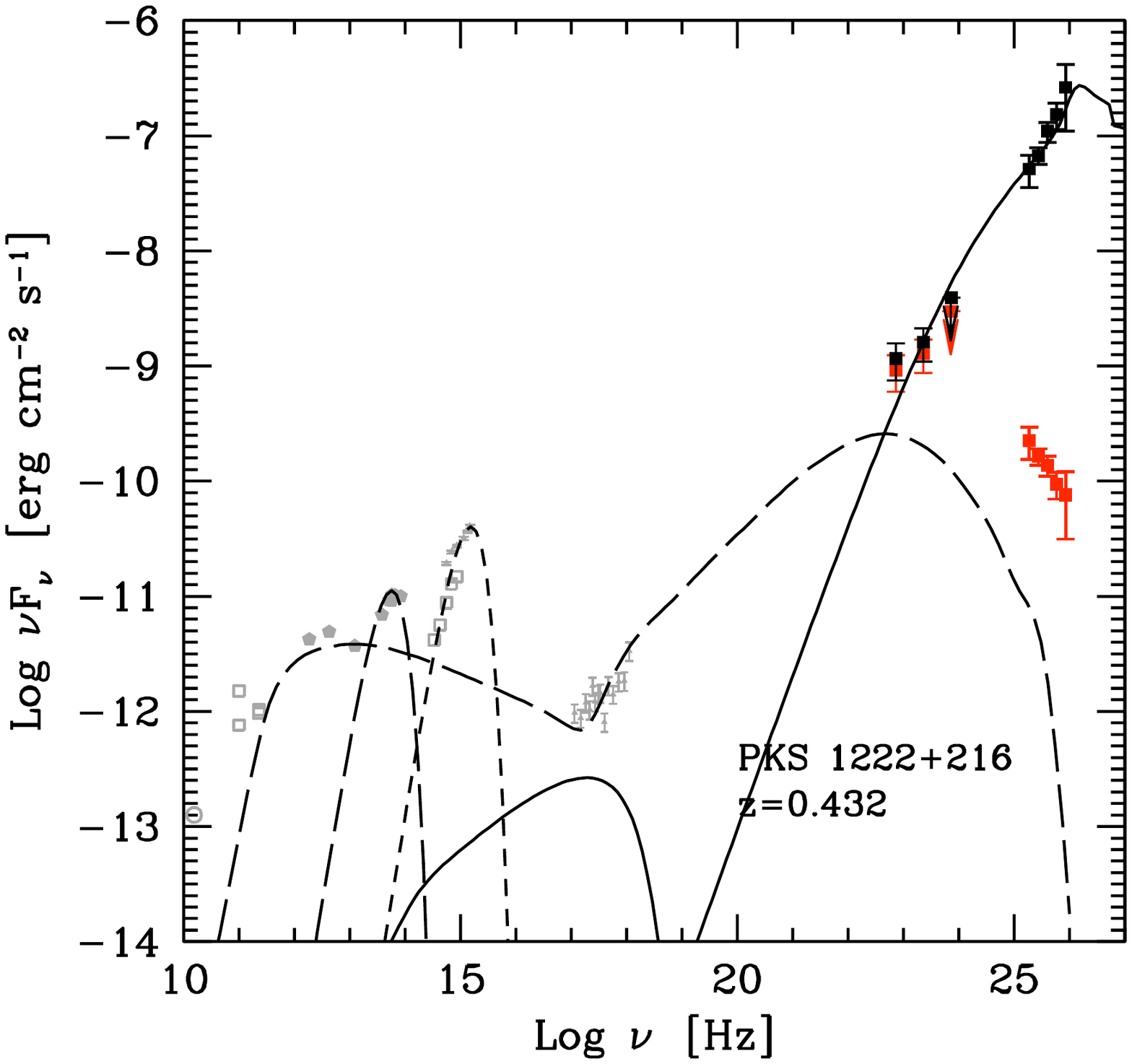}
\vspace{-2 truecm}
\caption{\label{fig:sedb1} Same as Fig.~\ref{fig:sedb} for the case $(B = 0.4 \, {\rm G}, M = 1.5 \cdot 10^{11} \, {\rm GeV})$ without photon-ALP oscillations in extragalactic space, but in addition the dashed and solid curves show the SED resulting from the considered two blobs which account for the $\gamma$-ray emission at high energy and VHE, respectively. }     
\end{figure}

\begin{figure}[h]
\centering
\includegraphics[width=.52\textwidth]{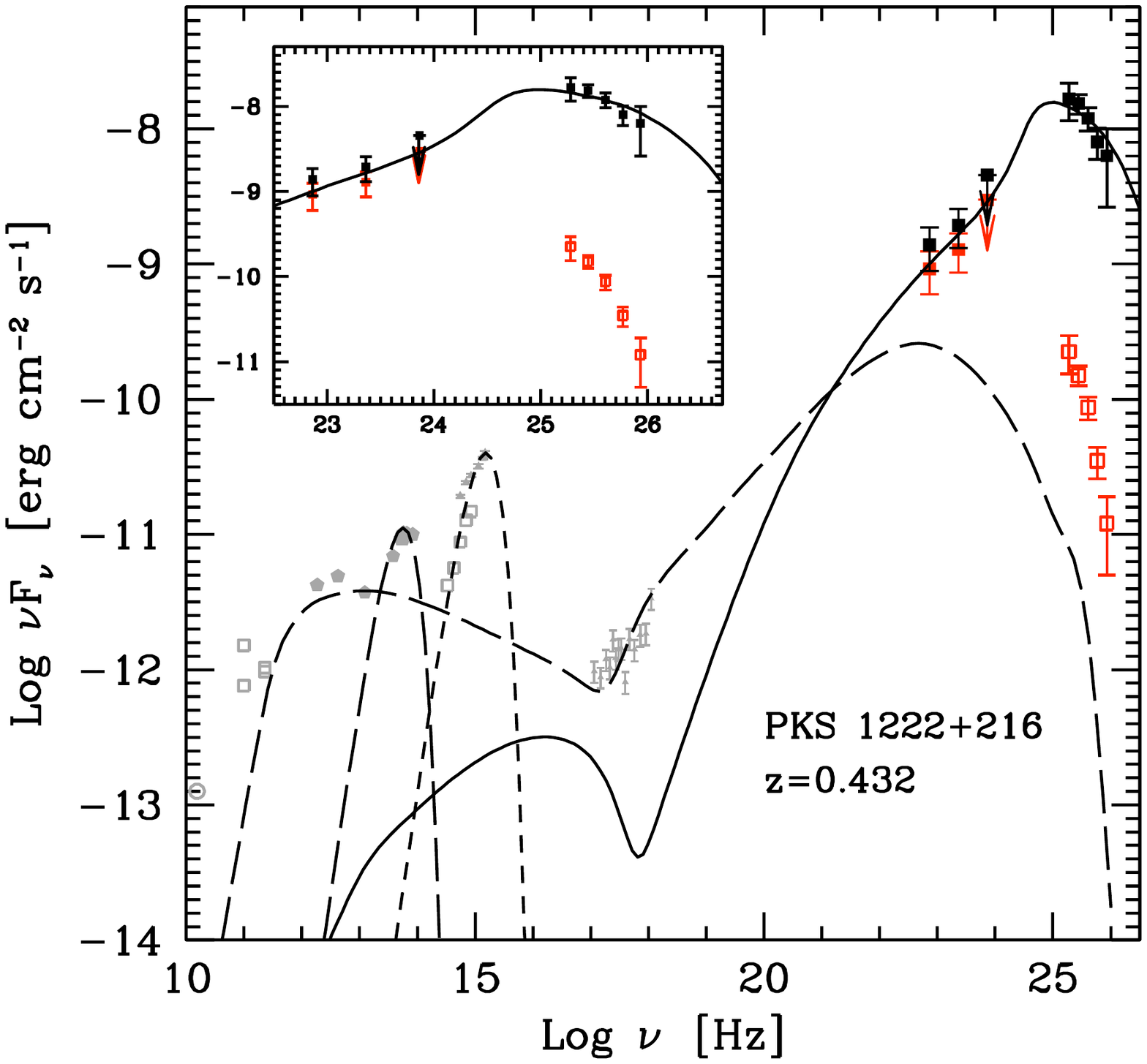}
\vspace{-2 truecm}
\caption{\label{fig:sedebl1} Same as Fig.~\ref{fig:sedebl} for the case $(B = 0.2 \, {\rm G}, M = 7 \cdot 10^{10} \, {\rm GeV})$ with photon-ALP oscillations in extragalactic space, but in addition the dashed and solid curves show the SED resulting from the considered two blobs which account for the $\gamma$-ray emission at high energy and VHE, respectively.}
\end{figure}

As already remarked, in an attempt to explain the observed MAGIC emission of PKS 1222+216 within conventional physics a particular model has been put forward~\cite{tavecchioetal2011}, which consists in a larger blob located {\it inside} the source responsible for the emission from IR to X-rays and a much smaller very compact blob accounting for the rapidly varying $\gamma$-ray emission detected by MAGIC. In order to avoid the BLR photon absorption, the smaller blob has been located {\it well outside} the BLR, namely at a large distance from the centre (we refer to the original paper~\cite{tavecchioetal2011} for a full discussion of the problems and a detailed description of this  model). Hence it looks natural to inquire whether a similar two-blob model can produce the SED needed in the present context -- namely to fit the black points in Figs. \ref{fig:seda}, \ref{fig:sedb} and \ref{fig:sedebl} -- with the key difference that now the smaller blob {\it lies close to the central engine}. Remarkably, this scenario works provided that the following parameters are chosen. Briefly, each region is specified by its size $r$, magnetic field $B$, bulk Lorentz factor $\Gamma$, electron normalization $K$, minimum, break and maximum Lorentz factors $\gamma _{\rm min}$, $\gamma _{\rm b}$, $\gamma _{\rm max}$ and slopes $n_1$, $n_2$. The electrons radiate through synchrotron and inverse Compton processes (considering both the internally produced synchrotron radiation and the external radiation of the BLR). For the larger region we use the same parameters of the original model while for the compact $\gamma$-ray blob region we have:

\begin{itemize}

\item Case ($B = 0.2 \, {\rm G}, M = 7 \cdot 10^{10} \, {\rm GeV}$) without photon-ALP oscillations in extragalactic space (see Fig. \ref{fig:seda1}) : $r = 2.2 \cdot 10^{14}$ cm, $B = 0.008 \, {\rm G}$, $\Gamma = 17.5$, $K = 6.2 \cdot 10^9$, $\gamma _{\rm min} =4 \cdot 10^3$, $\gamma _{\rm b} = 2.5 \cdot 10^5$, $\gamma _{\rm max}=4.9 \cdot 10^5$ and slopes $n_1 = 2.1$, $n_2 = 3.5$;

\item Case $(B = 0.4 \, {\rm G}, M = 1.5 \cdot 10^{11} \, {\rm GeV})$ without photon-ALP oscillations in extragalactic space (see Fig. \ref{fig:sedb1}): $r=2.2 \cdot 10^{14} \, {\rm cm}$, $B=0.0004 \, {\rm G}$, $\Gamma = 17.5$, $K = 2 \cdot 10^{11}$, $\gamma _{\rm min} = 5 \cdot 10^4$, $\gamma _{\rm b}  = 2.5 \cdot 10^6$, $\gamma _{\rm max}=4.9 \cdot 10^6$ and slopes $n_1 = 2.1$, $n_2 = 3.2$.

\item Case $(B = 0.2 \, {\rm G}, M = 7 \cdot 10^{10} \, {\rm GeV})$ with photon-ALP oscillations in extragalactic space (see Fig. \ref{fig:sedebl1}): $r = 2.2 \cdot 10^{14}$ cm, $B = 0.008 \, {\rm G}$, $\Gamma = 17.5$, $K = 6.7 \cdot 10^9$, $\gamma _{\rm min}=3\cdot 10^3$, $\gamma _{\rm b} = 1.2 \cdot 10^5$, $\gamma _{\rm max}=4.9 \cdot 10^5$ and slopes $n_1 = 2.1$, $n_2 = 3.5$.

\end{itemize}
Also here the relative position of the two regions is not relevant for the emission properties. 

The resulting SED is exhibited in Figs. \ref{fig:seda1}, \ref{fig:sedb1} for the cases ($B = 0.2 \, {\rm G}, M = 7 \cdot 10^{10} \, {\rm GeV}$) and $(B = 0.4 \, {\rm G}, M = 1.5 \cdot 10^{11} \, {\rm GeV})$ without photon-ALP oscillations in extragalactic space, respectively, and in Fig. \ref{fig:sedebl1} for the case ($B = 0.2 \, {\rm G}, M = 7 \cdot 10^{10} \, {\rm GeV}$) with photon-ALP oscillations in extragalactic space.


We find that an optimal choice to explain both {\it Fermi}/LAT and MAGIC observations corresponds to the case ($B = 0.2 \, {\rm G}, M = 7 \cdot 10^{10} \, {\rm GeV}$) without photon-ALP oscillations in extragalactic space, for which we obtain through Eq. (\ref{abc0606}) the black points shown in Fig. \ref{fig:seda1}. This result looks completely satisfactory, with the {\it Fermi}/LAT and MAGIC data well described by a high-energy bump peaking around $50 \, {\rm GeV}$ and a hight $\nu F_{\nu} \simeq 10^{- 8} \, {\rm erg} \, {\rm cm}^{- 2} \, {\rm s}^{- 1}$ corresponding to a luminosity $L_{\gamma} = 6 \cdot 10^{48} \, {\rm erg} \,  {\rm s}^{-1}$. On the other end, the case $(B = 0.4 \, {\rm G}, M = 1.5 \cdot 10^{11} \, {\rm GeV})$ looks satisfactory as far as the shape of the SED is concerned with the high-energy peak close to $500 \, {\rm GeV}$ as it can be seen from Fig. \ref{fig:sedb1}, but the implied luminosity of the $\gamma$-ray emission approaches  $L_{\gamma}=10^{51}$ erg s$^{-1}$ which appears unrealistic, since it is about 100 times larger than that of the most $\gamma$-ray bright blazars (see e.g.~\cite{gmt2009}). So, this result is unsatisfactory. We recall that the case $(B = 2 \, {\rm G}, M = 4 \cdot 10^{11} \, {\rm GeV})$ has already been ruled out due to the by far too high $\gamma$-ray peak. Finally, the inclusion of photon-ALP oscillations in extragalactic space makes the emitted spectrum slightly softer as it is evident from Fig. \ref{fig:sedebl1}, but the situation still remains completely satisfactory.

In conclusion, we find it a highly nontrivial circumstance that the benchmark case $(B = 0.2 \, {\rm G}$ and $M = 7 \cdot 10^{10} \, {\rm GeV})$ turns out to be the best one concerning both the efficiency for VHE photons to escape from the BLR and the SED of the particular two-blob model that we have adopted. Thus, it turns out to be by far our best option.

\bigskip
\bigskip
\bigskip

\section{CONCLUSIONS}

We have shown that the surprising $\gamma$-ray detection of PKS 1222+216 by MAGIC can be explained -- consistently with the simultaneous results by {\it Fermi}/LAT -- within a standard blazar model by adding the new assumption that inside the source photons can oscillate into ALPs. Our explanation assumes an average magnetic field with strength $B \simeq 0.2 \, {\rm G}$ in the jet up to the BLR and a value $M = 7 \cdot 10^{10} \, {\rm GeV}$ for the inverse coupling $\gamma\gamma a$. We remark that the emission model presented here is merely an example, and different and possibly more realistic scenarios can be constructed along similar lines. The main point we want to make is that with the photon-ALP oscillation mechanism at work the emission can well originate inside the BLR just like in conventional BL Lac models. As far as photon-ALP oscillations are concerned, their crucial role takes place in the source region before and just after the BLR. 

Needless to say, our scenario naturally applies also to the other FSRQs detected at VHE like 3C279 and PKS 1510-089~\cite{albertetal2008, aleksicetal2011b, wagnerbehera2010}, although these cases appear less problematic for the external emission scenario due to the absence of evident rapid ($t<1$ day) variability.

As already mentioned some alternative scenarios accounting for the puzzling features of PKS 1222+216 have been recently appeared in the literature. For instance, in~\cite{nalewajko2012} it is proposed that the VHE emission arises in the parsec-scale jet through the production of collimated beams of high-energy electrons by fast relativistic magnetic reconnection (see also~\cite{cerutti}). Alternatively, in~\cite{dermer,murase} it is assumed the existence of collimated beams of neutral particles produced in the inner jet through photo-meson reactions of ultra-high energy protons. Neutral particles can freely propagate to distances larger than the BLR and then produce ultra-relativistic leptons interacting with the IR radiation of the dusty torus. In turn, the collimated leptons would produce highly beamed synchrotron emission. If the magnetic field at pc scale is small enough, not exceeding 1 mG, the VHE synchrotron radiation preserves the rapid variability of the inner engine at the pc scale (but in this case the confinement of the clouds of the BLR looks problematic).

It appears to us remarkable that our proposed model lends itself to an observational test. Because photon-ALP oscillations can mitigate -- but not completely avoid -- the $\gamma$-ray absorption inside the BLR, a natural prediction is that at the optically-thin/optically thick transition around 30 GeV (in the source frame) the spectrum should display a feature. So, the absence of such a feature would be hard to explain in our model but would directly support scenarios in which the emission occurs outside the BLR as those discussed above.

We find it quite tantalizing that precisely the most favorable value $M = 7 \cdot 10^{10} \, {\rm GeV}$ for the effect considered in this paper corresponds to the most favorable case for a large-scale magnetic field of $B = 0.7 \, {\rm nG}$ in the DARMA scenario that enlarges the ``$\gamma$-ray horizon" and provides a natural solution to the cosmic opacity problem~\cite{dgr} (the DARMA scenario requires for the ALP mass $m < 1.7 \cdot 10^{- 10} \, {\rm eV}$ which is consistent with the present model). 

We want also to point out that it has recently been shown that ALPs with precisely the properties needed in our model naturally emerge in the Large Volume Scenario of IIB string compactifications~\cite{cicoli}. In addition, a very light ALP of the kind considered here is a viable candidate for the quintessential dark energy~\cite{carroll}. 

Very remarkably, an independent laboratory check of our proposal will be performed with the planned upgrade of the photon regeneration experiment ALPS at DESY~\cite{alps} and with the next generation of solar axion detectors like IAXO~\cite{iaxo}.

\section*{Acknowledgments}

M. R. thanks M. Meyer for his comments on an earlier version of the manuscript and F. T. thanks G. Ghisellini and G. Giovannini for useful discussions. It is really a pleasure to thank the unknown referee for help in clarifying the presentation.


\begin{thebibliography}{9}
\bibitem{commentA} 
We recall that discarding cosmological effects we have ${\tau}_{\rm eff} (E) = D/{\lambda}_{\rm eff} (E)$, where 
${\lambda}_{\rm eff} (E)$ is the effective photon mean free path and $D$ is the distance to the source.
\bibitem{comment31} 
They are independent of $\varphi_n$ because ${\cal M}_n (E_0, \varphi_n)$ arises from Eq. (\ref{t5L28q}) where $V(\psi)$ is an orthogonal matrix.
\end{thebibliography}
\end{document}